\documentclass[aps,prd,nofootinbib,twocolumn,showpacs,preprintnumbers,amsmath,amssymb,superscriptaddress,groupedaddress,floatfix]{revtex4-1}

\usepackage{amsmath,amssymb,bm,graphicx,epsfig,color,gensymb,feynmf}
\usepackage[caption=false]{subfig}

\newcommand{\be}{\begin{equation}}  
\newcommand{\ee}{\end{equation}}

\newcommand{\nl}{\nonumber \\ }

\allowdisplaybreaks

\begin{document}

\title{Deuterium target data for precision neutrino-nucleus cross sections}

\author{Aaron S. Meyer}
\email{asmeyer2012@uchicago.edu}
\affiliation{Enrico Fermi Institute and Department of Physics,  The University of Chicago, Chicago, Illinois, 60637, USA}
\affiliation{Fermi National Accelerator Laboratory, Batavia, Illinois 60510, USA}

\author{Minerba Betancourt}
\email{betan009@fnal.gov}
\affiliation{Fermi National Accelerator Laboratory, Batavia, Illinois 60510, USA}

\author{Richard Gran}
\email{rgran@d.umn.edu}
\affiliation{Department of Physics and Astronomy, University of Minnesota – Duluth, Duluth, Minnesota 55812, USA}

\author{Richard J. Hill}
\email{richardhill@uchicago.edu}
\affiliation{TRIUMF, 4004 Wesbrook Mall, Vancouver, British Columbia, V6T 2A3 Canada}
\affiliation{Perimeter Institute for Theoretical Physics, Waterloo, Ontario, N2L 2Y5 Canada}
\affiliation{Enrico Fermi Institute and Department of Physics,  The University of Chicago, Chicago, Illinois, 60637, USA}

\date{\today}

\begin{abstract}
  Amplitudes derived from scattering data on elementary targets are
  basic inputs to neutrino-nucleus cross section predictions.   A
  prominent example is the isovector axial nucleon form factor,
  $F_A(q^2)$, which controls charged current signal processes at
  accelerator-based neutrino oscillation experiments.  Previous
  extractions of $F_A$ from neutrino-deuteron scattering data rely on
  a dipole shape assumption that introduces an unquantified 
  error.   A new analysis of world data for neutrino-deuteron
  scattering is performed using a model-independent, and
  systematically improvable, representation of $F_A$.  A complete
  error budget for the nucleon isovector axial radius leads to
  $r_A^2=0.46(22) \,{\rm fm}^2$, with a much larger uncertainty than
  determined in the original analyses.   The quasielastic
  neutrino-neutron cross section is determined as  $\sigma(\nu_\mu n
  \to \mu^- p)\big|_{E_\nu =1\,{\rm GeV}} = 10.1(0.9) \times
  10^{-39}{\rm cm}^2$.  The propagation of nucleon-level constraints
  and uncertainties to nuclear cross sections is illustrated using
  MINERvA data and the GENIE event generator.   These techniques can
  be readily extended to other amplitudes and processes.   
\end{abstract}

\pacs{
  13.15.+g 
  14.60.Pq 
  14.20.Dh 
}

\maketitle{}

\section{Introduction} 

Current and next generation accelerator-based neutrino experiments are
poised to answer fundamental questions about
neutrinos~\cite{Acciarri:2015uup,Itow:2001ee,Adamson:2016tbq,Adamson:2016xxw,Chen:2007ae}.
Precise neutrino scattering cross sections on target nuclei are
critical to the success of these experiments.  These cross sections
are computed using nucleon-level amplitudes  combined with nuclear
models.  Determination of the requisite nuclear corrections presently
relies on data-driven
modeling~\cite{Andreopoulos:2009rq,Hayato:2009zz,Buss:2011mx,Golan:2012wx}
employing experimental
constraints~\cite{Drakoulakos:2004gn,AguilarArevalo:2010zc,AguilarArevalo:2010bm,AguilarArevalo:2010cx,Abe:2015awa,Anderson:2011ce,Rodrigues:2015hik,Megias:2016lke}.
{\it Ab initio} nuclear computations are beginning to provide additional
insight~\cite{Lovato:2013cua,Bacca:2014tla,Carlson:2014vla}.
Regardless of whether nuclear corrections are constrained
experimentally or derived from first principles, independent knowledge
of the elementary nucleon-level amplitudes is essential.  In this
paper, we address the problem of model-independent extraction of
elementary amplitudes from scattering data, and the propagation of
rigorous uncertainties through to nuclear observables.  
 
The axial-vector nucleon form factor, $F_A(q^2)$, is a prominent
source of uncertainty in any neutrino cross section program.   While
the techniques employed in the present paper may be similarly applied
to other elementary amplitudes, such as vector form
factors~\cite{vector}, we focus on the axial-vector form factor, which
is not probed directly in electron scattering measurements, and which
has large uncertainty.

The axial form factor is  constrained, with a varying degree of model
dependence, by neutron beta decay~\cite{Agashe:2014kda}, neutrino
scattering on nuclear targets heavier than
deuterium~\cite{Lyubushkin:2008pe,AguilarArevalo:2010zc,
  Brunner:1989kw, Pohl:1979zm, Auerbach:2002iy, Belikov:1983kg,
  Bonetti:1977cs}, pion electroproduction~\cite{Bernard:2001rs} and
muon capture~\cite{Andreev:2012fj}.  Existing data for the
neutrino-deuteron scattering process provide the most direct access to
the shape of the axial-vector nucleon form factor.  The assumption of
a neutron at rest and barely bound in the laboratory frame permits
unambiguous energy reconstruction, eliminating flux uncertainties.
The abundant neutrino scattering data on heavier targets involve
degenerate uncertainties from neutrino flux, and from large and 
model-dependent nuclear corrections, complicating the extraction of
nucleon-level amplitudes.  Antineutrino scattering on hydrogen would
entirely eliminate even the nuclear corrections required for
deuterium, but there are no high-statistics data for this process.
Given the importance of deuterium data for the axial form factor,  it
is imperative to quantify the constraints from existing data.

In this paper, we present the charged-current axial-vector nucleon
form factor and error budget determined from neutrino-deuterium
scattering data.  In place of the dipole assumption
(cf. Eq.~(\ref{eq:dipole}) below) used in previous analyses of the
form factor,  we employ the model-independent $z$ expansion%
\footnote{Formalism for $z$ expansion and nucleon form factors is
  described in  Refs.~\cite{Bhattacharya:2011ah,Hill:2010yb}, and
  several applications are  found in
  Refs.~\cite{Lorenz:2014vha,Epstein:2014zua,Lee:2015jqa,Bhattacharya:2015mpa}.
  Related formalism and applications may
  be found in~\cite{Hill:2006ub, Bourrely:1980gp,
    Boyd:1994tt,Boyd:1995sq,Lellouch:1995yv,Caprini:1997mu,Arnesen:2005ez,
    Becher:2005bg,Hill:2006bq,Bourrely:2008za,Bharucha:2010im,Amhis:2014hma,Bouchard:2013pna,Bailey:2015dka,Horgan:2013hoa,Lattice:2015tia,Detmold:2015aaa}.
}
parametrization.   The resulting uncertainty is significantly
larger than found in previous analyses~\cite{Bodek:2008epjc,
  Kuzmin:2007kr, Bernard:2001rs} of the deuterium constraint on the
axial form factor using multiple data sets.  This larger uncertainty
results from removing the dipole assumption, and from including
systematic errors for experimental acceptance corrections and for
model-dependent deuteron corrections.  The new constraints may be
readily implemented in nuclear models and neutrino event generators.

The remainder of the paper is structured as follows.  In
Sec.~\ref{sec:deut} we introduce the deuterium data sets and perform
fits to the dipole model for the axial form factor.  This is done in
order to compare with original publications, and to isolate the impact
of form factor shape assumptions versus other inputs or data
selections.  In Sec.~\ref{sec:z} we review the relevant $z$ expansion
formalism, and redo fits from Sec.~\ref{sec:deut} replacing dipole
with $z$ expansion.  Several features of these fits indicate
potentially underestimated systematic errors in corrections that were
applied to data in the original publications.  Section~\ref{sec:syst}
describes a range of systematics tests.  We consider several sources
of systematic errors in more detail in Sec.~\ref{sec:exptsyst}, and
redo fits in Sec.~\ref{sec:extract}, where we present final results
for $F_A(q^2)$.   In Sec.~\ref{sec:nuclear} we illustrate the
propagation of errors to several derived observables, including the
isovector axial nucleon radius and total neutrino-nucleon quasielastic
cross sections.  The incorporation of nucleon-level uncertainties in
nuclear cross sections is illustrated with MINERvA
data~\cite{Fiorentini:2013ezn}.  Section~\ref{sec:conclusion} provides
a summary and conclusion. 

\section{Deuterium data and dipole fits \label{sec:deut}}

\begin{table*}[t]
  \caption{\label{tab:constants}
    Inputs from the original publications, BNL1981~\cite{Baker:1981su},
    ANL1982~\cite{Miller:1982qi} and FNAL1983~\cite{Kitagaki:1983px}, 
    and our default inputs.  See text for details. 
  }
  \begin{ruledtabular}
    \begin{tabular}{cccccc}
      Input & BNL1981 & ANL1982 & FNAL1983 & This work & Reference \\
      \hline
      $g_A = F_A(0)$ & -1.23 & -1.23 & -1.23 & -1.2723 & \cite{Agashe:2014kda}
      \\
      $\mu_p - \mu_n-1$ & 3.708 & 3.71 & 3.708 & 3.7058 & \cite{Agashe:2014kda}
      \\
      $F_{Vi}$ & Olsson~\cite{Olsson:1978dw} & Olsson~\cite{Olsson:1978dw} & Olsson~\cite{Olsson:1978dw} & BBA2005 & \cite{Bradford:2006yz}
      \\
      $F_P$ & PCAC & PCAC & PCAC & PCAC  & (\ref{eq:PCAC})
      \\
      Deuteron correction & Singh~\cite{Singh:1971md} & Singh~\cite{Singh:1971md}
      & Singh~\cite{Singh:1971md} &
      Singh & \cite{Singh:1971md}
      \\
      lepton mass & $m_\mu =m_\mu$ except ABC & $m_\mu =m_\mu$   & $m_\mu =m_\mu$ except ABC & $m_\mu =m_\mu$
      \\
      $Q^2$ range & $0.06-3 \,{\rm GeV}^2$ & $0.05-2.5\,{\rm GeV}^2$ & $0-3\,{\rm GeV}^2$ 
      \\
      $N_{\rm bins}$ & 49 & 49 & 30  
      \\
      $N_{\rm events}$ & 1236 & 1792 & 354 
      \\
      kinematic cut & $Q^2 \ge 0.06\,{\rm GeV}^2$ & $Q^2 \ge 0.05\,{\rm GeV}^2$ & $Q^2 \ge 0.10\,{\rm GeV}^2$ &
    \end{tabular}
  \end{ruledtabular}
\end{table*}

The world data from deuterium bubble chamber experiments consists of
deuterium fills of the ANL 12-foot deuterium bubble chamber
experiment~\cite{Mann:1973pr,Barish:1977qk,Miller:1982qi}, the BNL
7-foot deuterium bubble chamber experiment~\cite{Baker:1981su}, and
the FNAL 15-foot deuterium bubble chamber
experiment~\cite{Kitagaki:1983px}.   We refer below to these
experiments as ANL1982, BNL1981 and FNAL1983, respectively.%
\footnote{ An updated BNL data set was presented in
  Ref.~\cite{Kitagaki:1990vs} with a factor $\approx 2$ increase in
  number of events.  However, we were unable to extract a sufficiently
  precise $Q^2$ distribution of events from this reference, since the
  data were presented on a logarithmic scale
  (cf. Ref.~\cite{Kitagaki:1990vs}, Fig.~5).  We thus consider only
  the events from the BNL1981 data set.}

\subsection{Fits to $Q^2$ distributions}

Extracting the axial form factor from data requires information about
all other aspects of the scattering cross-section.  The original
publications used a variety of different inputs for axial ($g_A$) and
magnetic ($\mu_p-\mu_n$) couplings, vector and pseudoscalar form
factors, nuclear corrections, and muon mass corrections.
Table~\ref{tab:constants} displays the input choices made in the
original publications for each of the three considered data sets, as
well as the updated inputs used for the remainder of this paper.%
\footnote{ Form factor notations and conventions are as in
  Ref.~\cite{Bhattacharya:2011ah}.}

The vector form factors are constrained by invoking isospin symmetry
and constraints of electron-nucleon scattering data.  In place of the
Olsson vector form factors~\cite{Olsson:1978dw},  we use the so-called
BBA2005 parametrization that is commonly employed in contemporary
neutrino studies~\cite{Bradford:2006yz}.   Similar results were
obtained using the BBA2003~\cite{Budd:2003wb} and
BBBA2007~\cite{Bodek:2007ym} parametrizations.
Recent developments, connected with the so-called ``proton radius
puzzle'', point to potential shortcomings in previous extractions of
the vector form
factors~\cite{Pohl:2010zza,Bernauer:2013tpr,Lee:2015jqa}.  A
systematic study of the vector form factors similar to the $z$
expansion analysis of the axial form factor presented here is
undertaken in Refs.~\cite{Lee:2015jqa,vector}.

For the pseudoscalar form factor $F_P$, we employ the partially
conserved axial current (PCAC) ansatz,
\begin{align}\label{eq:PCAC}
F_P^{\rm PCAC}(q^2) = {2 m_N^2 F_A(q^2) \over m_\pi^2 - q^2} \, .
\end{align}
The free-nucleon form factors $F_A$ and $F_P$ are functions of the
four momentum transfer $q^2$ from the lepton to the nucleon, and
$m_N=0.9389\,{\rm GeV}$, $m_\pi=0.14\,{\rm GeV}$ are the masses of the
nucleon and the pion.  The effects of the pseudoscalar form factor are
suppressed in the limit of small lepton mass, and its uncertainties
are negligible in most applications involving accelerator neutrino
beams, including this analysis. 

Nuclear corrections relating the free neutron cross section,
$d\sigma^{n}$, to the deuteron cross section, $d\sigma^D$,  may be
parametrized as
\begin{align}\label{eq:nucl}
  {d\sigma^{D}\over dQ^2} = R(Q^2,E_\nu) {d\sigma^{n}\over dQ^2} \,,
\end{align}
where $d\sigma^{D}/dQ^2$ denotes the deuteron differential cross
section with respect to the intrinsically positive $Q^2$ = $-q^2$.%
\footnote{ For definiteness in the deuteron case, we let $Q^2$ in
  Eq.~(\ref{eq:nucl}) denote the leptonic momentum transfer.   This
  definition is consistent with  the experimental reconstruction,
  which assumed the kinematics for scattering from a free neutron in
  the presence of a spectator proton carrying opposite momentum to the
  neutron.}
The model of Ref.~\cite{Singh:1971md} was used in the original
analyses, with  $R(Q^2,E_\nu) \approx R(Q^2)$ independent of neutrino
energy, and $R(Q^2) \to 1$ above $Q^2\approx 0.2\,{\rm GeV}^2$.  We
retain this model as default, but examine deviations from this simple
description below in Sec.~\ref{sec:syst}, using the calculations of
Ref.~\cite{Shen:2012xz}.

The neutrino-neutron quasielastic cross section may be written in a
standard form
\begin{align}\label{eq:CCQE}
{d\sigma^n \over dQ^2} \propto {1\over E_\nu^2} \bigg[ A(Q^2) \mp
  B(Q^2){ s-u \over m_N^2}+ C(Q^2) {(s-u)^2 \over m_N^4} \bigg] \,,
\end{align}
where $s-u = 4 E_\nu m_N - Q^2 - m_\mu^2$ is the difference of
Mandelstam variables, $A$, $B$ and $C$ are quadratic functions of
nucleon form factors~\cite{LlewellynSmith:1971zm},  and the
vector-axial interference term $B$ changes sign for the $\bar{\nu}p$
scattering process.  In the BNL1981 and FNAL1983 data sets, the lepton
mass was neglected inside the functions $A(Q^2)$, $B(Q^2)$ and
$C(Q^2)$ of Eq.~(\ref{eq:CCQE}), but retained in other kinematic
prefactors.  In our analysis, we retain the complete lepton mass
dependence.  

The event distributions in $Q^2$ have been obtained by digitizing the
relevant plots from the original publications.
Table~\ref{tab:constants} gives the $Q^2$ range and bin size, the
total number of events,%
\footnote{ For BNL1981 and ANL1982, the digitized number of events in
  each $Q^2$ bin was rounded to the nearest integer, resulting in the
  same total numbers, 1236 and 1792 respectively, quoted in the
  original publications.  For FNAL1983, the digitization produced
  near-integer results in each $Q^2$ bin, but the total summed event
  number, 354, differs from the value 362 quoted in the original
  publication.    } and the minimum $Q^2$ retained in the original
analyses.  In each case, events in a lowest $Q^2$ bin were omitted
from fits, and only FNAL1983 reports these events.   We retain the same
binning and minimum $Q^2$ cut in our default fits. These distributions
are included as Supplemental Material to the present paper~\cite{self}.

\subsection{$E_\nu$ distributions and flux\label{sec:Enu}}

An advantage of the $\nu_\mu d \rightarrow \mu^{-} p p$ process in an
exquisite device like a bubble chamber is the accurate reconstruction
of the neutrino energy for each event.  Cross section parameters can
be constrained from the $Q^2$ distribution despite poorly controlled
uncertainties in {\it ab initio} neutrino flux estimates.  This is
especially valuable for the low energy  ANL1982 and BNL1981 data,
whose neutrino energy spectrum significantly influences the shape of
the $dN/dQ^2$ distribution through the energy-dependent kinematic
limit corresponding to a backscattered lepton.

Unfortunately, event-level kinematics from the deuterium data sets are
no longer available and unbinned likelihood fits using  the $E_\nu$
and $Q^2$ dependence of the cross section cannot be repeated.
However, the one-dimensional distribution of events in reconstructed
neutrino energy, $dN/dE_\nu$, may be extracted from the original
publications, and we use this information to reconstruct the flux
self-consistently.  This subsection describes the procedure we use,
including some subtle points required for later interpretation of the
form factor fits.

The differential neutrino flux is determined by 
\begin{align}\label{eq:flux}
{d \Phi(E_\nu) \over dE_\nu} \propto {1\over \sigma^n(E_\nu,F_A)}
{dN^n\over dE_\nu} \,,  
\end{align}
where $\sigma^n(E_\nu,F_A)$ is the free-neutron quasielastic cross
section, and $dN^n/dE_\nu$ is the energy distribution of free-neutron
events that would be obtained in the experimental flux.  The constant
of proportionality in Eq.~(\ref{eq:flux}) is determined by the number
of target deuterons and the time duration of the experiment.    Let us
normalize the energy distribution according to 
\begin{align}\label{eq:norm}
\int_0^\infty dE_\nu {dN^n\over dE_\nu} = {\cal N} \int_{Q^2_{\rm
    min}}^\infty dQ^2 {dN^D \over dQ^2} \,. 
\end{align}
Consistency in Eq.~(\ref{eq:norm}) is obtained when ${\cal N} =
\hat{\cal N}(F_A,Q^2_{\rm min},R)$, where 
\begin{align}\label{eq:NQ2}
  \hat{\cal N}(F_A,Q^2_{\rm min},R) = \dfrac{ \int_0^{\infty}  dQ^2
    \dfrac{dN^n}{dQ^2}  } { \int_{Q^2_{\rm min}}^\infty dQ^2
    \dfrac{dN^D}{dQ^2} } \,.
\end{align}
The right-hand side of Eq.~(\ref{eq:NQ2}) may be computed using a
given $dN^n/dE_\nu$, and depends on  $R(Q^2,E_\nu)$ and $F_A(q^2)$
through Eqs.~(\ref{eq:nucl}) and (\ref{eq:flux}).  Using the flux from
Eq.~(\ref{eq:flux}), we have finally,
\begin{multline}\label{eq:theory}
   \left( {dN^D\over dQ^2} \right)^{\rm theory} \\ = {\cal N}_{\rm
     fit} \int_0^\infty dE_\nu \dfrac{ R(Q^2,E_\nu)
     \dfrac{d\sigma^n}{dQ^2}(E_\nu, F_A, Q^2) } { \sigma^n(E_\nu,F_A)}
        {dN^n\over dE_\nu} \,,
\end{multline}
where a fit parameter, ${\cal N}_{\rm fit}$, has been introduced for
the normalization.

Choosing ${\cal N}= \hat{\cal N}(F_A, Q^2_{\rm min}, R)$ in
Eq.~(\ref{eq:norm}) would correspond to ${\cal N}_{\rm fit}=1$.  In
order to avoid the explicit computation of the integrals
(\ref{eq:NQ2}),  we instead take ${\cal N}=1$, corresponding to the
expectation ${\cal N}_{\rm fit} = {\cal N}(F_A, Q^2_{\rm min}, R)$.
We allow the parameter ${\cal N}_{\rm fit}$ to float unconstrained in
the fits, with an independent parameter for each experiment.  

We emphasize that $dN^n/dE_\nu$ in Eq.~(\ref{eq:flux}) represents the
energy distribution of free-neutron events that would be obtained in
the experimental flux; this distribution  is obtained from the energy
distribution of observed events in deuterium by correcting for nuclear
effects, for events lost due to the $Q^2_{\rm min}$ cut,  and for
other experimental effects.  Such corrections were applied to the
energy distribution presented in the BNL1981 data set, but not in the ANL1982
and FNAL1983 data sets.   The effect of applying or not applying these
corrections is found to be small, as discussed below in
Sec.~\ref{sec:flux}.

For later comparison, we compute the ratios (\ref{eq:NQ2}) with a
nominal dipole axial form factor ($m_A=1\,{\rm GeV}$,
cf. Eq.~(\ref{eq:dipole}) below), neglecting deuteron corrections
($R=1$),  and at a nominal $E_\nu=1\,{\rm GeV}$ neutrino energy, for
the $Q^2_{\rm min}$ values employed in the BNL1981, ANL1982, FNAL1983
data sets:%
\footnote{ While $E_\nu=1\,{\rm GeV}$ is close to the peak energy for
  the BNL1981 and ANL1982 data sets, the FNAL1983 data set involved
  higher energy.  However, these ratios have mild energy dependence
  above $E_\nu \sim 1\,{\rm GeV}$, e.g. at $E_\nu=10\,{\rm GeV}$ the
  result is $\hat{\cal N}(Q^2_{\rm min}=0.10\,{\rm GeV}^2) \approx
  1.25$.  }
\begin{align}\label{eq:Ns}
  \hat{\cal N}(Q^2_{\rm min}=0.06\,{\rm GeV}^2)&\approx 1.13, \nl
  \hat{\cal N}(Q^2_{\rm min}=0.05\,{\rm GeV}^2)&\approx 1.11, \nl
  \hat{\cal N}(Q^2_{\rm min}=0.10\,{\rm GeV}^2)&\approx 1.23.
\end{align}
We expect these numbers to be approximately reproduced in
$\mathcal{N}_{\rm fit}$ when the deviation from $\hat{N}=1$ in
Eq.~(\ref{eq:NQ2}) is dominated by the $Q^2_{\rm min}$ cut.

Two further complications result in technical subtlety but do not
affect the fit results.  First, the binned event rate $dN/dE_\nu$ for
ANL1982 is provided in a prior publication \cite{Barish:1978pj} that
used a subset of about half the events.  A second complication is the
finite bin width of the $dN/dE_\nu$ distributions, which would yield
unphysical discontinuities when displaying ANL and BNL $dN/dQ^2$
spectra at best fit.  This effect is the result of convoluting a low
energy flux with a differential cross section that has an
energy-dependent kinematic limit.  We use an interpolation algorithm
to produce smoothed fluxes with 500 bins in energy over the original
range of data.  Nearly identical fit results are obtained regardless
of whether the interpolation is a cubic spline, linear, or whether the
original binning is used, so this step is primarily cosmetic.
The smoothed and unsmoothed $E_\nu$ distributions are included as
Supplemental Material to the present paper~\cite{self}.

\subsection{Dipole fits}

\begin{table}[t]
  \caption{\label{tab:mAdipole}
    Dipole axial mass extracted in original publications, our extraction
    using parameter inputs as in the original publications,
    and our extraction using updated constants and vector form factors as in Table~\ref{tab:constants}.
    Errors are statistical only.  
  }
    \begin{ruledtabular}
    \begin{tabular}{llll}
      & $m_A^{\rm dipole}({\rm ref})$ & $m_A^{\rm dipole}({\rm old})$ & $m_A^{\rm dipole}({\rm new})$ \\
      \hline
BNL 1981 \cite{Baker:1981su} & 1.07(6) & 1.07(5) & 1.05(5) \\
ANL 1982 \cite{Miller:1982qi} & 1.05(5) & 1.05(5) & 1.02(5) \\
FNAL 1983 \cite{Kitagaki:1983px} & $1.05^{+0.12}_{-0.16}$ & 1.20(11) & 1.17(10)  
    \end{tabular}
  \end{ruledtabular}
\end{table}

Our results for the axial form factor will differ from  the analyses
in the original publications.   These differences arise from a number
of sources:  updated numerical inputs in Table~\ref{tab:constants};
not using unbinned likelihood fits; and differences in axial form
factor shape assumptions.  In order to understand these differences,
we begin by restricting attention to the dipole ansatz,
\begin{align}\label{eq:dipole}
F^{\rm dipole}_A(q^2) = F_A(0) \left( 1 - {q^2\over m_A^2}
\right)^{-2} \,, 
\end{align}
and compare to fits in the original publications.

Table~\ref{tab:mAdipole} gives results for fits to the dipole ansatz
(\ref{eq:dipole})  for the axial form factor.   The table shows
``flux-independent'' results from the original experiments, which
performed unbinned likelihood fits to event-level data.  Our results
are from a Poisson likelihood fit to the binned $Q^2$ distribution of
events obtained with a neutrino flux given by smoothing the binned
reconstructed neutrino energy distribution (divided by theoretical
cross section), as described in Sec.~\ref{sec:Enu}.  Fits to the
binned log-likelihood function are found by minimizing the function
\begin{align}
-2 {\rm log}[\mathcal{L}\left(F_A\right)] =  2 \sum_i \left[
  \mu_i(F_A) - n_i + n_i {\rm
    log}\left(\frac{n_i}{\mu_i(F_A)}\right)\right] \, ,
\end{align}
where $n_i$ is the number of events in the $i$th bin, and $\mu_i$ is
the theory prediction (\ref{eq:theory}) for the bin.  Errors
correspond to changes of $1.0$ in the $-2$LL function.

Because we do not use an unbinned likelihood fit,  we do not expect
precise agreement even when the original choices of constants in
Table~\ref{tab:constants} are used.   Comparing the first two columns
of Table~\ref{tab:mAdipole},  the size of the resulting statistical
uncertainties are approximately equal, and only FNAL shows a discrepancy
in central value.   A similar exercise was
performed in Refs.~\cite{Budd:2003wb,Bodek:2003ed,Budd:2004bp}, and
similar results were obtained.  Having reproduced the original
analyses to the extent possible, we will proceed with the updated
constants as in the final column of  Table~\ref{tab:constants}.

\section{$z$ expansion analysis \label{sec:z} }

The dipole assumption (\ref{eq:dipole}) on the axial form factor shape
represents an unquantified systematic error.   We now remove this
assumption, enforcing only the known analytic structure that the form
factor inherits from QCD.   We investigate the constraints from
deuterium data in this more general framework.   A similar analysis
may be performed using future lattice QCD calculations in place of
deuterium data. 

\subsection{$z$ expansion formalism}

The axial form factor obeys the dispersion relation,
\begin{align}
F_A(q^2) = {1\over \pi} \int_{t_{\rm cut}}^\infty dt^\prime {{\rm Im}
  F_A(t^\prime + i0) \over t^\prime - q^2} \,,
\end{align}
where  $t_{\rm cut}= 9m_\pi^2$ represents the leading three-pion
threshold for states that can be produced by the axial current.  The
presence of singularities along the positive real axis implies that a
simple Taylor expansion of the form factor in the variable $q^2$  does
not converge for $|q^2| \ge 9 m_\pi^2 \approx 0.18\,{\rm GeV}^2$.
Consider the new variable obtained by mapping the domain of
analyticity onto the unit circle~\cite{Bhattacharya:2011ah}, 
\begin{align}\label{eq:zdef}
z(q^2,t_{\rm cut}, t_0) = { \sqrt{t_{\rm cut} - q^2} - \sqrt{t_{\rm
      cut} - t_0} \over \sqrt{t_{\rm cut} - q^2} + \sqrt{t_{\rm cut} -
    t_0} } \,, 
\end{align}
where $t_0$, with $-\infty < t_0 < t_{\rm cut}$, is an arbitrary
number that may be chosen for convenience.  In terms of the new
variable we may write a convergent expansion,
\begin{align}\label{eq:FAz}
F_A(q^2) = \sum_{k=0}^{k_{\rm max}} a_k z(q^2)^k \,, 
\end{align}
where the expansion coefficients $a_k$ are dimensionless numbers
encoding nucleon structure information.  

\begin{table}[t]
  \caption{\label{tab:t0}
    Maximum value of $|z|$ for different $Q^2$ ranges and choices of $t_0$. 
    $t_0^{\rm optimal}$ is defined in Eq.~(\ref{eq:t0opt}). 
  }
    \begin{ruledtabular}
    \begin{tabular}{ccc} 
      $Q^2_{\rm max}\, [{\rm GeV}^2]$ & $t_0$ & $|z|_{\rm max}$ 
      \\
      \hline
      1.0 & 0 & 0.44 \\
      3.0 & 0 & 0.62 \\
      1.0 & $t_0^{\rm optimal}(1.0\,{\rm GeV}^2)=-0.28\,{\rm GeV}^2$ & 0.23 \\
      3.0 & $t_0^{\rm optimal}(1.0\,{\rm GeV}^2)=-0.28\,{\rm GeV}^2$ & 0.45 \\
      3.0 & $t_0^{\rm optimal}(3.0\,{\rm GeV}^2)=-0.57\,{\rm GeV}^2$ & 0.35 \\
    \end{tabular}
  \end{ruledtabular}
\end{table}

In any given experiment, the finite range of $Q^2$ implies a maximal
range for $|z|$ that is less than unity. We denote by $t_0^{\rm
  optimal}(Q^2_{\rm max})$ the choice which minimizes the maximum size
of $|z|$ in the range $-Q^2_{\rm max} \le q^2 \le 0$.  Explicitly,
\begin{align}\label{eq:t0opt}
t_0^{\rm optimal}(Q^2) = t_{\rm cut} ( 1 - \sqrt{1 + Q^2_{\rm
    max}/t_{\rm cut}} ) \,. 
\end{align}
Table~\ref{tab:t0} displays $|z|_{\rm max}$ for several choices of
$Q^2_{\rm max}$ and $t_0$.  

The choice of $t_0$ can be optimized for various applications.  We
have in mind applications with data concentrated below $Q^2= 1\,{\rm
  GeV}^2$,  and therefore take as default choice, 
\begin{align} \label{eq:t0choice}
\bar{t}_0 = t_0^{\rm optimal}(1\,{\rm GeV}^2) \approx -0.28 \,{\rm
  GeV}^2 \,, 
\end{align}
minimizing the number of parameters that are necessary to describe
data in  this region.  Inspection of Table~\ref{tab:t0} shows that the
form factor expressed as $F_A(z)$ becomes approximately linear.  For
example, taking $|z|_{\rm max}=0.23$ implies that quadratic, cubic,
and quartic terms enter at the level of $\sim 5\,\%$, $1\,\%$ and
$0.3\,\%$. 

The asymptotic scaling prediction from perturbative
QCD~\cite{Lepage:1980fj}, $F_A \sim Q^{-4}$, implies the series of
four sum rules~\cite{Lee:2015jqa}
\begin{align}\label{eq:sumrules}
\sum_{k=n}^\infty k(k-1)\cdots (k-n+1) a_{k} = 0 \,, \quad n=0,1,2,3
\,.
\end{align}
We enforce the sum rules (\ref{eq:sumrules}) on the coefficients,
ensuring that the form factor falls smoothly to zero at large $Q^2$.
Together with the $Q^2=0$ constraint, this leaves
$N_a = k_{\rm max}-4$ free parameters in Eq.~(\ref{eq:FAz}). 
From Eq.~(\ref{eq:sumrules}), it can be shown~\cite{Lee:2015jqa} that
the coefficients behave as $a_k \sim k^{-4}$ at large $k$.  We remark
that the dipole ansatz (\ref{eq:dipole}) implies the coefficient
scaling law $|a_k| \sim k$ at large $k$, in conflict with perturbative
QCD.

In addition to the sum rules, an examination of explicit spectral
functions and scattering data~\cite{Bhattacharya:2011ah} motivates the bound
of
\begin{align} \label{eq:bound1}
|a_k/a_0| \le 5.
\end{align}
As noted above, from Eq.~(\ref{eq:sumrules}), the coefficients behave
as $a_k \sim k^{-4}$ at large $k$.  We invoke a falloff of the
coefficients at higher order in $k$,
\begin{align} \label{eq:bound2}
|a_k/a_0|\le 25/k \,, \quad  k>5.
\end{align}
The bounds  are enforced with a Gaussian penalty on the coefficients
entering the fit.  We investigate fits using a range of $k_{\rm max}$,
other choices of $t_0$, and alternatives to Eqs.~(\ref{eq:bound1}) and
(\ref{eq:bound2}), which are briefly reported in Sec.~\ref{sec:syst}.

\subsection{$z$ expansion basic fit results \label{sec:zfits}}

\begin{table*}[t]
  \caption{\label{tab:zfit}
    Fits to $z$ expansion using the same data and constants as the final column of
    Table~\ref{tab:mAdipole}.  
    ``LL'' denotes log likelihood. Errors on $z$ expansion determinations of $r_A^2$
    are determined from the error matrix, all others correspond to $\Delta(-2{\rm LL})=1$.  
    $N_a=k_{\rm max}-4$ denotes the number of free expansion
    coefficients in the $z$ expansion fit (\ref{eq:FAz}) with scheme choice (\ref{eq:t0choice}), sum rule constraints (\ref{eq:sumrules}),
    and bounds (\ref{eq:bound1}), (\ref{eq:bound2}).
    The final column is the number of bins, including bins with zero data.
    For $N_a=4$ the resulting fit parameters are displayed in Eq.~(\ref{eq:ak}). 
  }
  \begin{ruledtabular}
    \begin{tabular}{r|ccc|ccc|ccc|ccc|c}
      && Dipole &&& $N_a=3$ &&& $N_a=4$ &&& $N_a=5$ & \\
      \hline
      Experiment  &
      $-2$LL & ${\cal N}_{\rm fit}$ & $r_A^2\,[{\rm fm}^2]$ &
      $-2$LL & ${\cal N}_{\rm fit}$ & $r_A^2\,[{\rm fm}^2]$ &
      $-2$LL & ${\cal N}_{\rm fit}$ & $r_A^2\,[{\rm fm}^2]$ &
      $-2$LL & ${\cal N}_{\rm fit}$ & $r_A^2\,[{\rm fm}^2]$ &
      $N_{\rm bins}$ 
      \\ 
\hline
      BNL1981  & 70.9 & $1.14^{+0.08}_{-0.07}$ & $0.424(44)$
               & 76.1 & $1.14^{+0.12}_{-0.11}$ & $0.36(21)$ 
               & 73.4 & $1.13^{+0.13}_{-0.11}$ & $0.25(21)$
               & 71.0 & $1.13^{+0.13}_{-0.12}$ & $0.18(21)$ & 49 \\
      ANL1982  & 58.6 & $1.15^{+0.06}_{-0.06}$ & $0.444(44)$ 
               & 62.3 & $1.15^{+0.10}_{-0.09}$ & $0.38(19)$ 
               & 60.9 & $1.14^{+0.10}_{-0.10}$ & $0.31(19)$
               & 59.9 & $1.14^{+0.11}_{-0.10}$ & $0.27(19)$ & 49 \\ 
      FNAL1983 & 38.2 & $1.17^{+0.16}_{-0.13}$ & $0.337(61)$
               & 39.1 & $1.21^{+0.24}_{-0.20}$ & $0.61(28)$ 
               & 39.1 & $1.21^{+0.25}_{-0.21}$ & $0.60(28)$ 
               & 39.1 & $1.20^{+0.26}_{-0.21}$ & $0.58(32)$ & 29 \\
    \end{tabular}
  \end{ruledtabular}
\end{table*}

\begin{figure}[h!]
  \begin{center}
    \includegraphics[width=0.48\textwidth]{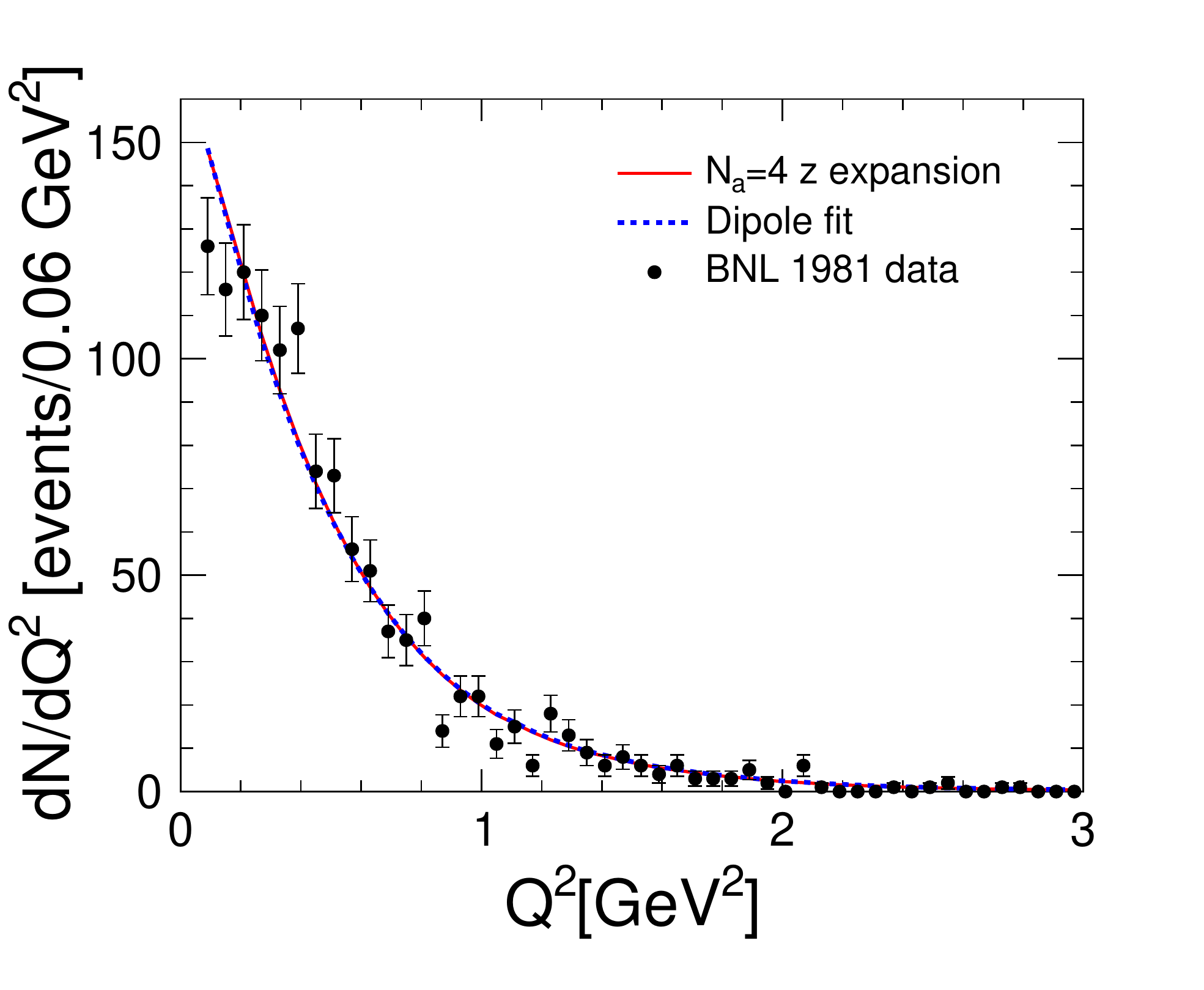}\vspace{-4mm}
    \includegraphics[width=0.48\textwidth]{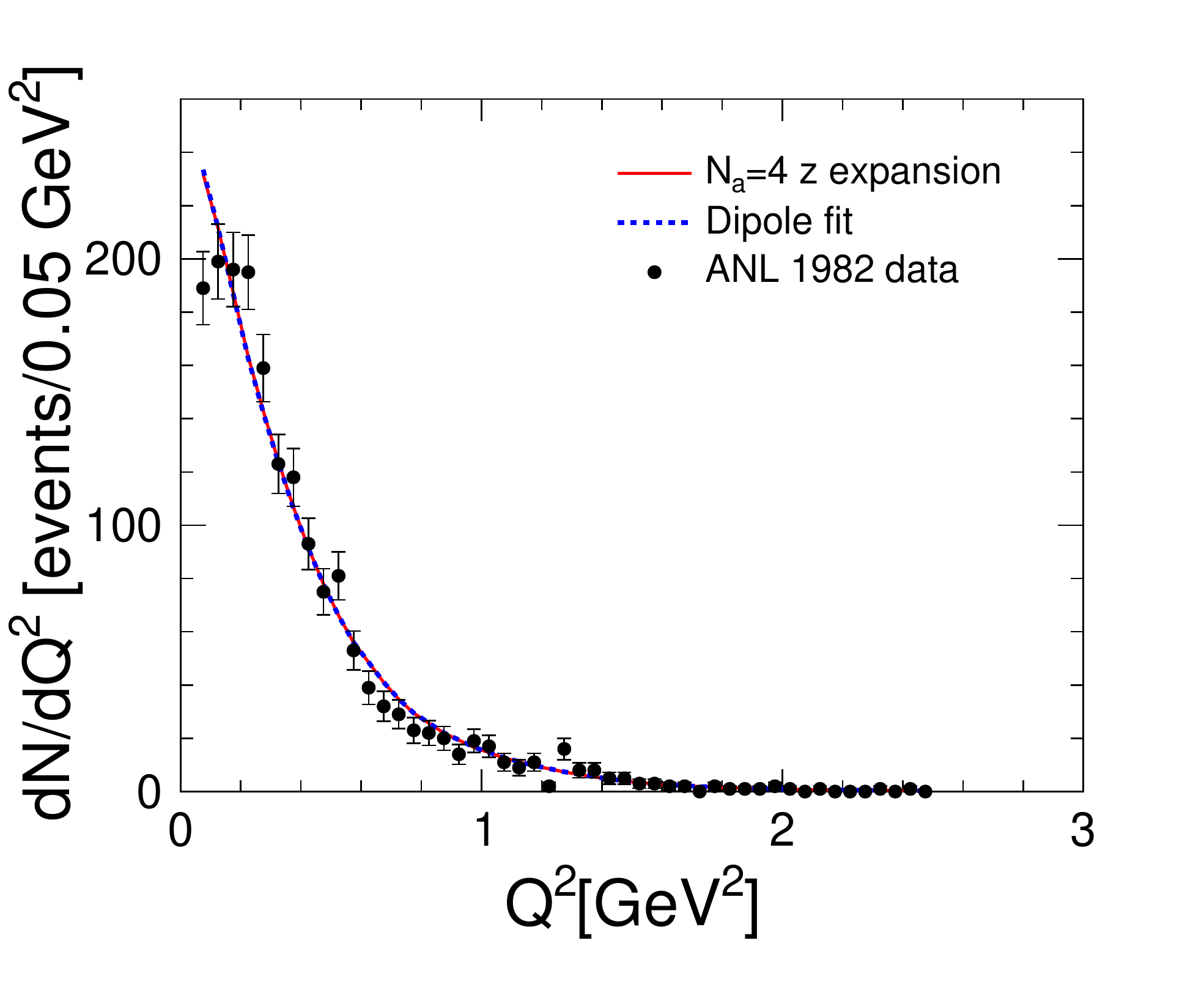}\vspace{-4mm}    
    \includegraphics[width=0.48\textwidth]{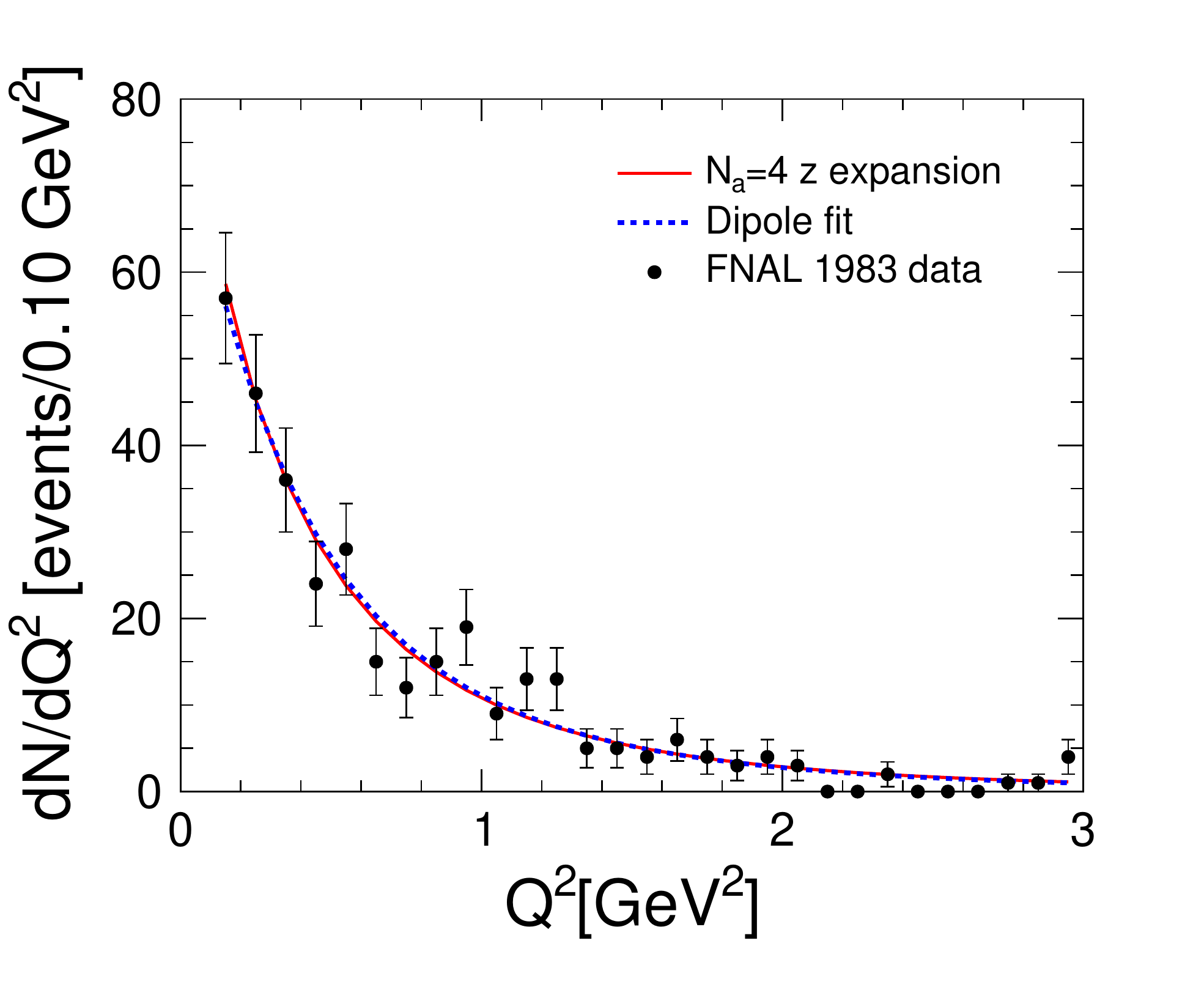}
    \caption{\label{fig:Q2}
    Experimental data and best fit curves corresponding to dipole and $N_a=4$ $z$ expansion in
    Table~\ref{tab:zfit}, for BNL1981 (top pane), ANL1982 (middle pane) and FNAL1983 (bottom pane). 
  }
  \end{center}
\end{figure}

\begin{figure}[h!]
    \includegraphics[width=0.48\textwidth]{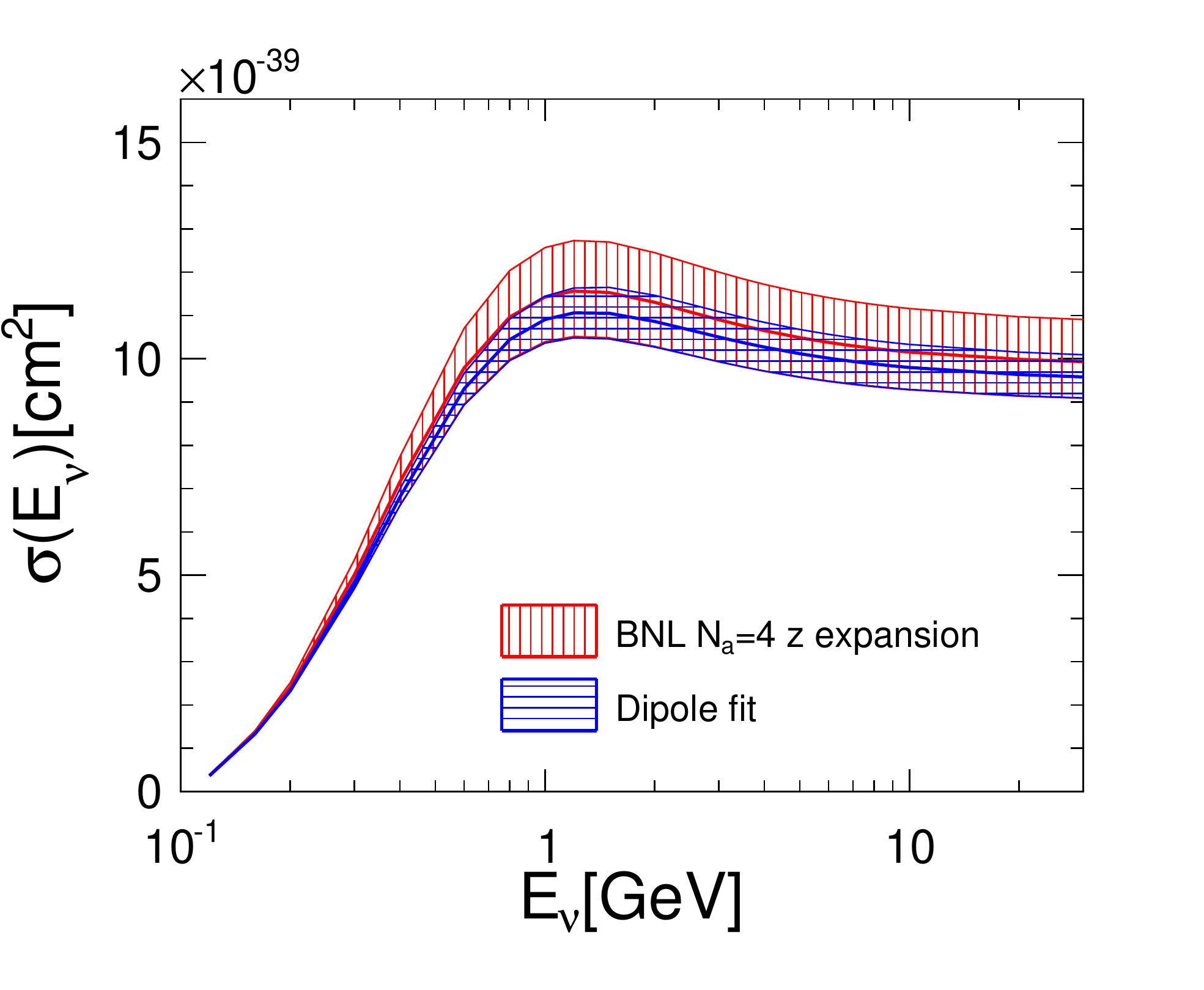}\vspace{-5mm}
    \includegraphics[width=0.48\textwidth]{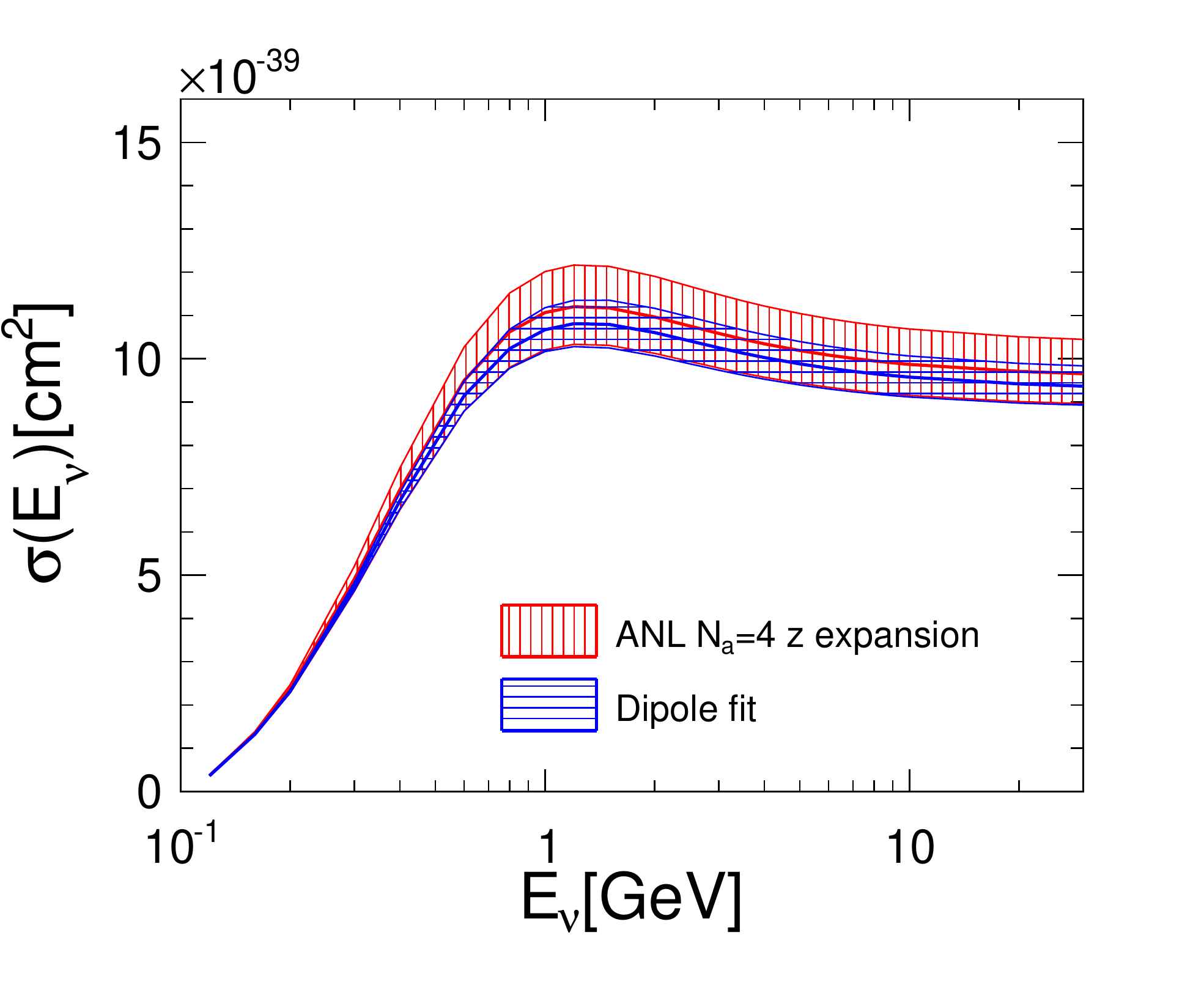}\vspace{-5mm}    
    \includegraphics[width=0.48\textwidth]{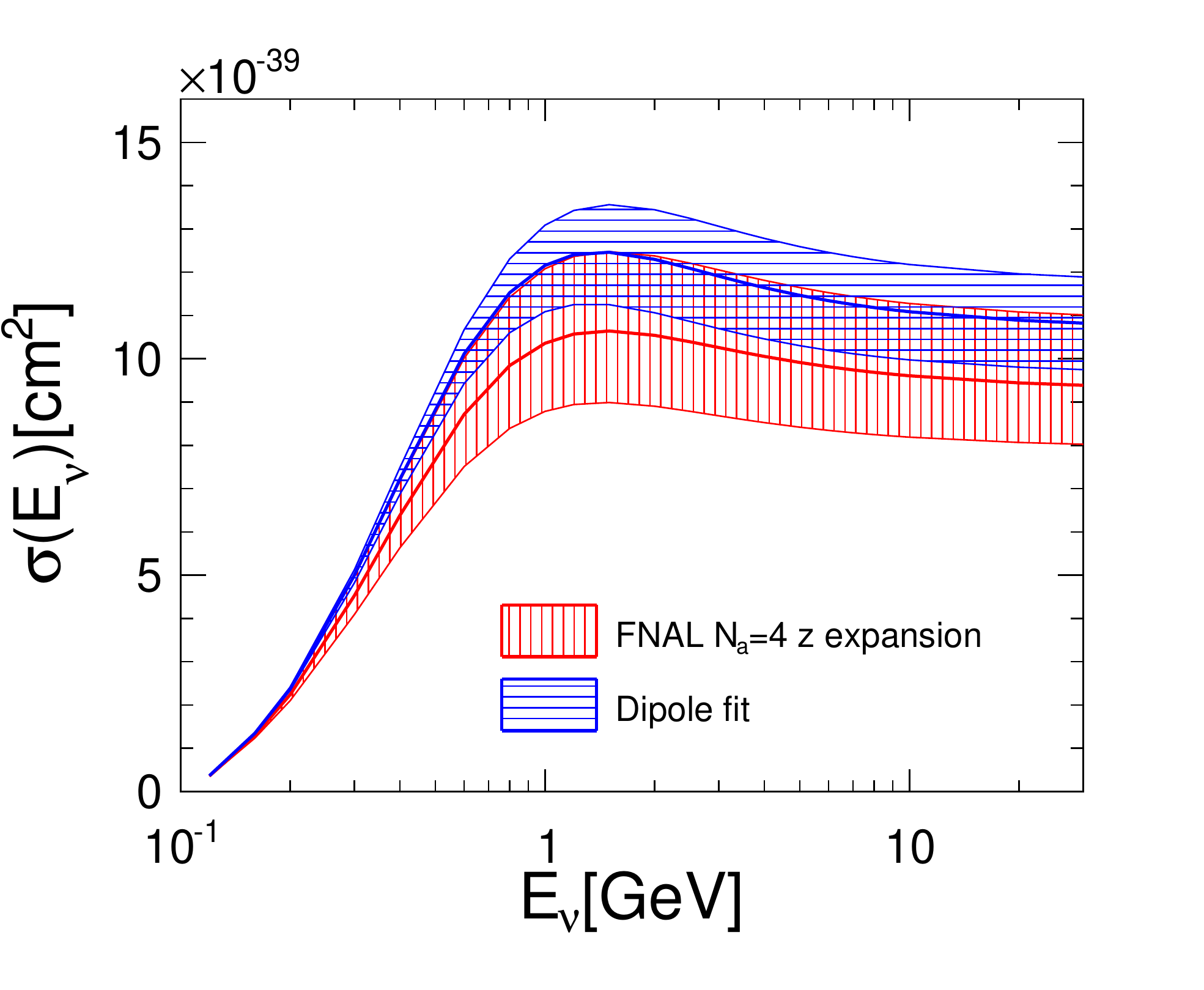}
  \caption{\label{fig:sigma3GeV}
    Best fit curves and errors propagated from deuterium to free-neutron cross section,
    for BNL1981 (top pane), ANL1982 (middle pane) and FNAL1983 (bottom pane). 
    Blue (horizontal stripes) corresponds to dipole and red (vertical stripes)
    to $N_a=4$ $z$ expansion in
    Table~\ref{tab:zfit}.
  }
\end{figure}

Using the same data sets and constants as described in
Sec.~\ref{sec:deut} and summarized in Table~\ref{tab:constants}, we
perform fits replacing dipole axial form factor with $z$ expansion as
in Eq.~(\ref{eq:FAz}).   We use the scheme choice (\ref{eq:t0choice}),
enforce the sum rule constraints (\ref{eq:sumrules}), and use the
default bounds on the coefficients $a_k$ in
Eqs.~(\ref{eq:bound1}), (\ref{eq:bound2}).   The results are
summarized in Table~\ref{tab:zfit} and displayed in Figs.~\ref{fig:Q2}
and~ \ref{fig:sigma3GeV}.  The coefficients corresponding to the fits with 
$N_a=4$ free parameters in Table~\ref{tab:zfit} are 
\begin{align}\label{eq:ak}
  &[a_1, a_2, a_3, a_4 ] \nl &=\left\{ \begin{array}{lr} $[2.24(10),\,
      0.6(1.0),\, -5.4(2.4), 2.2(2.7)]$ & ({\rm BNL})\\ $[2.25(10),\,
      0.2(0.9),\, -4.9(2.3), 2.7(2.7)]$ & ({\rm ANL})\\ $[2.02(14),\,
      -1.2(1.5),\, -0.7(2.9), 0.1(2.8)]$ & ({\rm FNAL})\\
  \end{array}
  \right. \,, 
\end{align}
where (symmetrized) errors correspond to a change of 1.0 in the $-2$LL
function. 

Table~\ref{tab:zfit} summarizes $z$ expansion fits with different
numbers of free parameters.  Focusing on the first order coefficient, 
\begin{align}\label{eq:kmax}
  &[a_1({\rm BNL}), \, a_1({\rm ANL}),\, a_1({\rm FNAL})] \nl
  &=\left\{
  \begin{array}{lr}  
    $[2.23(10),\, 2.23(10),\, 2.02(14) ]$\,, & N_a=3 \\  $[2.24(10),\,
      2.25(10),\, 2.02(14) ]$\,, & N_a=4 \\ $[2.22(10),\, 2.25(10),\,
      2.02(14) ]$\,, & N_a=5 
  \end{array}
  \right. \,.
\end{align}
As discussed after Eq.~(\ref{eq:t0choice}), $z^2$, $z^3$, $z^4$, etc.,
terms in the $z$ expansion become increasingly irrelevant,
corresponding to $|z|_{\rm max} \ll 1$ in Table~\ref{tab:t0}.  This is
borne out by the data, which determines a form factor with
coefficients in Eq.~(\ref{eq:ak}) of order 1.0 that mostly do not push
the Gaussian bounds, and a leading coefficient in Eq.~(\ref{eq:kmax})
that is approximately the same regardless of whether terms beyond
order $z^3$ are included. 

The axial ``charge'' radius is defined via the form factor slope at
$q^2=0$,
\begin{align}\label{eq:rA}
  {1\over F_A(0)} {dF_A \over dq^2}\bigg|_{q^2=0} \equiv \frac16 r_A^2
  \,.
\end{align}
For a general scheme choice $t_0\ne 0$, this quantity depends on all
the coefficients in the $z$ expansion.  Table~\ref{tab:zfit}
illustrates that $r_A$ is poorly constrained without the restrictive
dipole assumption.  We will provide a final value for the axial radius
from deuterium data after discussion of systematic errors in the next
section.

The normalization factor ${\cal N}_{\rm fit}$ is also included in
Table~\ref{tab:zfit}.  This parameter is allowed to float without
bounds, but returns values consistent with the approximation (\ref{eq:Ns})
to the expectation (\ref{eq:NQ2}).

\begin{figure}[tb]
\begin{center}
\includegraphics[width=0.49\textwidth]{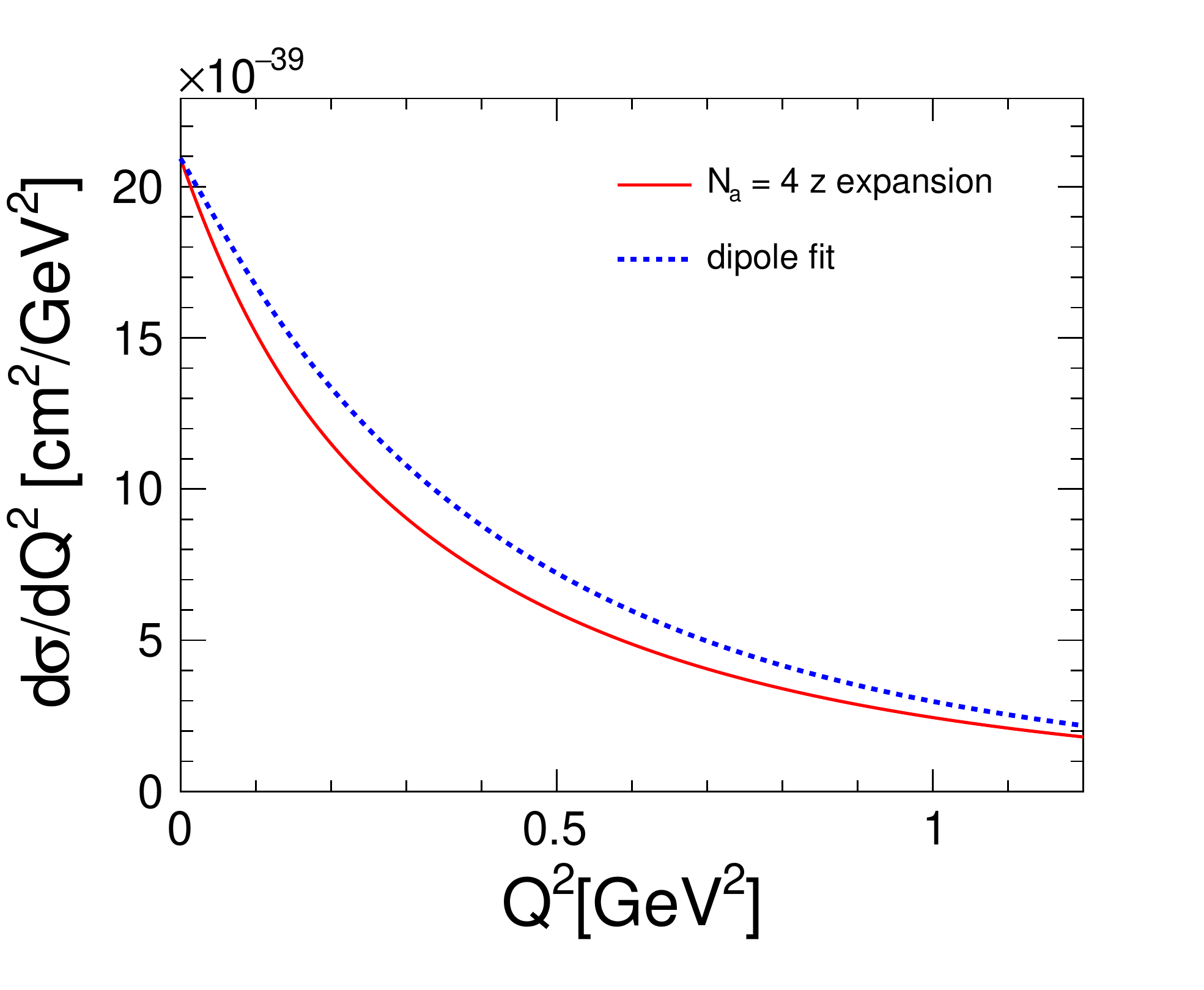}
\caption{
\label{fig:dsigma_compare}
Absolutely normalized $d\sigma^n/dQ^2$ at $E_\nu=10~{\rm GeV}$ for dipole (blue)
and $z$-expansion axial form factor central values
as in the FNAL1983 results of Figs.~\ref{fig:Q2} and Fig.~\ref{fig:sigma3GeV}.
}
\end{center}
\end{figure}

An interesting feature of the fits displayed in Fig.~\ref{fig:Q2} is
that whereas the best-fit $dN/dQ^2$ curves for dipole and $z$
expansion are very similar in the considered $Q^2$ range, derived
observables such as the radius in Table~\ref{tab:zfit}, and the
absolutely normalized cross section in Fig.~\ref{fig:sigma3GeV}, can
be markedly different.  The presence of the $Q^2_{\rm min}$ cut, and
the lack of an absolutely normalized flux, explains this situation,
which is most apparent for FNAL1983.  To illustrate,
Fig.~\ref{fig:dsigma_compare} shows the absolutely normalized
$d\sigma^n/dQ^2$ computed using the central value dipole and
$z$-expansion axial form factors for FNAL1983 in Figs.~\ref{fig:Q2}
and~\ref{fig:sigma3GeV}.  Omitting the lowest-$Q^2$ data, and
applying an overall normalization factor obscures the difference
between these curves. 

The normalization parameter ${\cal N}_{\rm fit}$ appearing in
Eq.~(\ref{eq:theory})  is not externally constrained in our shape
fits.  The uncertainty after fitting yields $\sim \pm 10\%$ for
BNL1981 and ANL1982 and $\sim \pm 20\%$ for FNAL1983, which is
significantly larger than the $\sim 3\%$ to $\sim 5\%$ uncertainty
from Poisson statistics.  A simple Poisson constraint would not be
adequate considering uncertainties from acceptance and deuterium
corrections described later.   A rate+shape fit with a correctly
motivated uncertainty on ${\cal N}_{\rm fit}$ could in principle
produce a somewhat better constrained form factor and cross section.

\subsection{Residuals analysis \label{sec:res}}

The best fits are still a relatively poor description of the data,
apparent in both Table~\ref{tab:zfit} and Fig.~\ref{fig:Q2}.   This
was briefly discussed in the thesis~\cite{Miller:1981thesis} that
accompanies the ANL1982 publication: the theoretical curve is too high
at very low $Q^2$, becoming too low above 0.2 GeV$^2$,  and too high
again around 0.7 GeV$^2$.  Similarly, the BNL1981 publication
discusses the possibility of residual scanning biases with a kinematic
dependence  that mimics evidence for second-class currents violating
the symmetries of QCD (cf. Ref.~\cite{Baker:1981su}, Fig.~5).
These observations motivate a careful examination of systematic
uncertainties assigned in the fits. 

\begin{figure}[tb]
\begin{center}
\includegraphics[width=0.49\textwidth]{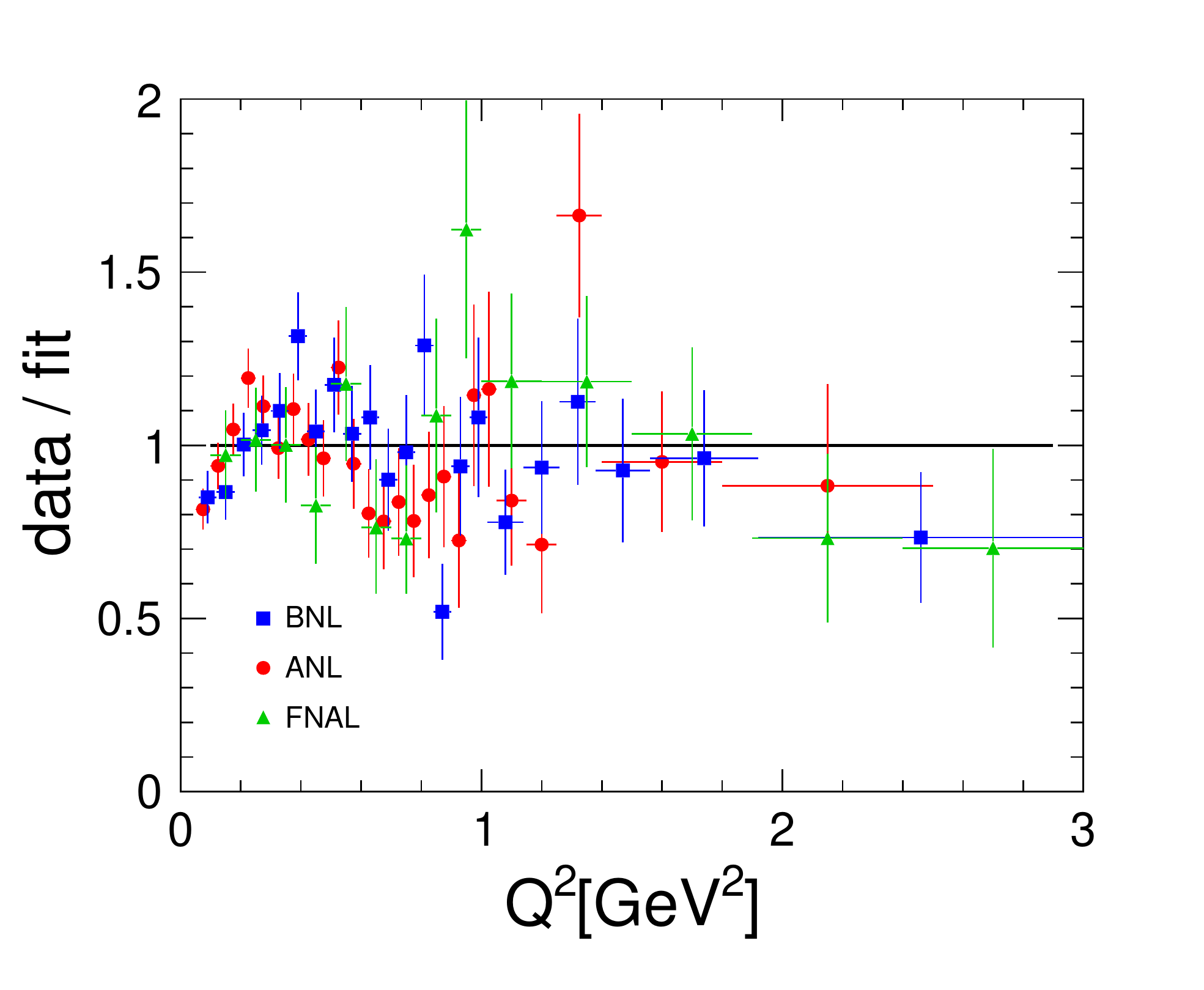}
\caption{
\label{fig:residual}
Data divided by best fit prediction for the $Q^2$ distributions
displayed in Fig.~\ref{fig:Q2}, for BNL(blue) ANL (red), and FNAL
(green). Calculated $\chi^2/N_{\text{bins}}$ are 35.3/22, 41.2/25,
and 10.7/14 for BNL, ANL, and FNAL respectively.}
\end{center}
\end{figure}

The preference of the experiments for a common $Q^2$-dependent
distortion can be illustrated by comparing the residual discrepancy
between the data and the best fit curves from Fig.~\ref{fig:Q2} in a
single plot, shown in Fig.~\ref{fig:residual}.%
\footnote{ For definiteness, the best fit curve is from a simultaneous
  fit to the BNL, ANL and FNAL data sets.  A nearly identical plot is
  obtained if different best fit curves for each data set are used.  }
The distortion at lowest $Q^2$ is clearly significant.  The data also seem
to agree on potential distortions in the range $0.25 < Q^2 < 3.0$.
However the null hypothesis, that the data in this range were drawn from a
flat distribution, yields P-value of 0.12 and is not exceptional.  In
order to use a $\chi^2$ fit for this P-value and to improve the plot
readability, the upper bins in each data set were combined.

Form factors described by the $z$ expansion, hence any form factor
consistent with QCD, cannot 
accommodate such localized distortions of the $Q^2$ spectrum (the dipole
ansatz similarly cannot accommodate such distortions).
The $R(Q^2)$ model for deuterium
used by the original experiments asymptotically approaches unity and
also does not cause such distortions.  It is interesting to consider whether
more complete deuteron correction models, such 
as Ref.~\cite{Shen:2012xz} (considered below in Fig.~\ref{fig:nuclear}), 
could produce such distortions.%
\footnote{ 
  Calculations of multinucleon effects for heavier nuclei like
  carbon exhibit qualitatively similar characteristics
  throughout this region of $Q^2$~\cite{Martini:2009uj, Gran:2013kda}.}
Finally, the impact of residual scanning biases should also be accounted for.
In the next sections we turn to the question of assigning a systematic
uncertainty to account for such effects.  

\section{Systematic tests \label{sec:syst}} 

Fits using different choices for constructing the $z$ expansion form
factors should yield equivalent results for physical observables: a
dependence on such choices would indicate an underestimated systematic
uncertainty.  Similarly, fits using different ranges of $Q^2$ should
yield equivalent results.

\subsection{Form factor scheme dependence}

A test with variations of the number of free parameters was presented
in Eq.~(\ref{eq:kmax}) of the previous section.  In order to translate
other test fits into parameters that can be compared side-by-side,
we will consider in all cases the dimensionless shape parameter
defined by
\begin{align} \label{eq:shape}
\bar{a}_1 \equiv a_1|_{t_0=\bar{t}_0} \equiv -4 (t_{\rm cut}
-\bar{t}_0) F_A^\prime(\bar{t}_0) \,,
\end{align}
where $\bar{t}_0 \equiv t_0^{\rm optimal}(1\,{\rm GeV}^2)=-0.28\,{\rm
  GeV}^2$, as in Eq.~(\ref{eq:t0choice}).  To motivate the choice
(\ref{eq:shape}), note that since $z$ is a small parameter, the form
factor is approximately linear when expressed as a function of $z$.
The slope of this approximately linear function is the essential shape
parameter determined by the data, and for convenience we define the
slope at $q^2=\bar{t}_0$.  [The axial radius is similarly defined as
  the form factor slope at $q^2=0$ in Eq.~(\ref{eq:rA}).]

\subsubsection{Magnitude of bound}

Consider first the numerical value of the bound (\ref{eq:bound1}).
For definiteness, we impose a coefficient falloff, $a_k \sim 1/k$, as
in Eq.~(\ref{eq:bound2}).  Focusing on $N_a=4$,
\begin{align}
&[\bar{a}_1({\rm BNL}), \, \bar{a}_1({\rm ANL}),\, \bar{a}_1({\rm
     FNAL})] \nl &= \left\{\!\!
  \begin{array}{ll}
    $[2.18(8),\,\,\,2.17(8),\,\,\,2.01(12)]$, &
    \!\left|\dfrac{a_k}{a_0}\right| \le {\rm
      min}\left(3,\dfrac{15}{k}\right)
    \\ \\ $[2.23(10),\,2.25(10),\,2.02(14)]$, &
    \!\left|\dfrac{a_k}{a_0}\right| \le {\rm
      min}\left(5,\dfrac{25}{k}\right)
    \\ \\ $[2.36(15),\,2.41(15),\,2.02(17)]$, &
    \!\left|\dfrac{a_k}{a_0}\right| \le {\rm
      min}\left(10,\dfrac{50}{k}\right)
  \end{array}\right.
  . 
\end{align}
Results are consistent within errors.  The very conservative bound
$|a_k/a_0|\le 10$ would lead to an error that is $\sim 50\%$ larger
than our default $|a_k/a_0|\le 5$.  

\subsubsection{Choice of $t_0$} 

Next, consider the choice of $t_0$.%
\footnote{For $t_0=t_0^{\rm optimal}(1\,{\rm GeV}^2)=-0.28\,{\rm
    GeV}^2$, by design, the shape parameter is identified with the
  linear coefficient of the $z$ expansion in Eq.~(\ref{eq:FAz}).  Since
  $\bar{a}_1$ [Eq.~(\ref{eq:shape})] is a physical observable, it can be
  computed for any choice of $t_0\ne \bar{t}_0$.}
A different choice of $t_0$ requires more parameters to achieve the
same truncation error, $\sim |z|^{N_a+1}$.  We compare the default
case of $t_0=-0.28\,{\rm GeV}^2$ and $N_a=4$ to the case of $t_0=0$
and $N_a=7$,%
\footnote{ Both cases have $|z|_{\rm max}^{N_a+1} \approx 0.02$ in the
  range $0<Q^2<3\,{\rm GeV}^2$.  } finding 
\begin{align}
  &[\bar{a}_1({\rm BNL}), \, \bar{a}_1({\rm ANL}),\, \bar{a}_1({\rm
       FNAL})] \nl &= \left\{
  \begin{array}{ll}
    $[2.24(10),\,2.25(10),\,2.02(14)]$ & ( N_a=4,\, t_0=\bar{t}_0 )
    \\ $[2.22(9),\,2.21(10),\,2.02(14)]$ & ( N_a=7,\, t_0=0 )
  \end{array}\right. \,, 
\end{align}
where the errors are propagated using the covariance matrix for the
coefficients $a_k$.   Nearly identical results are obtained for
different choices of $t_0$.

\subsection{Subsets of the $Q^2$ range \label{sec:range}} 

A nonstatistical scatter of data points about the best fit curves is
apparent in Fig.~\ref{fig:Q2}, and indicated by the poor fit quality
in Table~\ref{tab:zfit}.  Removing subsets of the data at high or low
$Q^2$ will help isolate sources of tension between data and fit. 

\begin{table*}[t]
  \caption{\label{tab:zfit1}
    Same as Table~\ref{tab:zfit}, but fitting only to data with $Q^2 \le 1\,{\rm GeV}^2$. 
    For $N_a=4$ the resulting fit parameters are displayed in Eq.~(\ref{eq:ak1}). 
  }
  \begin{ruledtabular}
    \begin{tabular}{r|ccc|ccc|ccc|ccc|c}
      && Dipole &&& $N_a=3$ &&& $N_a=4$ &&& $N_a=5$ & \\
      \hline
      Experiment  &
      $-2$LL & ${\cal N}_{\rm fit}$ & $r_A^2\,[{\rm fm}^2]$ &
      $-2$LL & ${\cal N}_{\rm fit}$ & $r_A^2\,[{\rm fm}^2]$ &
      $-2$LL & ${\cal N}_{\rm fit}$ & $r_A^2\,[{\rm fm}^2]$ &
      $-2$LL & ${\cal N}_{\rm fit}$ & $r_A^2\,[{\rm fm}^2]$ &
      $N_{\rm bins}$ 
      \\ 
\hline
      BNL1981  & 24.7 & $1.16^{+0.08}_{-0.08}$ & $0.348(48)$
               & 27.2 & $1.17^{+0.14}_{-0.13}$ & $0.32(22)$ 
               & 27.0 & $1.17^{+0.14}_{-0.13}$ & $0.28(22)$ 
               & 26.6 & $1.16^{+0.14}_{-0.13}$ & $0.24(22)$ & 16 \\
      ANL1982  & 28.2 & $1.14^{+0.07}_{-0.06}$ & $0.452(52)$
               & 31.7 & $1.15^{+0.10}_{-0.09}$ & $0.38(19)$ 
               & 30.5 & $1.14^{+0.10}_{-0.10}$ & $0.31(20)$ 
               & 29.2 & $1.13^{+0.11}_{-0.10}$ & $0.24(20)$ & 19 \\
      FNAL1983 & 8.3  & $1.16^{+0.26}_{-0.18}$ & $0.33(12)$  
               & 8.3  & $1.22^{+0.29}_{-0.23}$ & $0.54(31)$ 
               & 8.2  & $1.23^{+0.29}_{-0.24}$ & $0.56(29)$ 
               & 8.1  & $1.24^{+0.30}_{-0.24}$ & $0.57(26)$ & 9 
    \end{tabular}
  \end{ruledtabular}
\end{table*}

\begin{table*}[t]
  \caption{\label{tab:zfit3}
    Same as Table~\ref{tab:zfit}, but fitting only to data with $Q^2 \ge 0.2\,{\rm GeV}^2$.
    For $N_a=4$ the resulting fit parameters are displayed in Eq.~(\ref{eq:ak3}). 
  }
  \begin{ruledtabular}
    \begin{tabular}{r|ccc|ccc|ccc|ccc|c}
      && Dipole &&& $N_a=3$ &&& $N_a=4$ &&& $N_a=5$ && \\
      \hline
      Experiment  &
      $-2$LL & ${\cal N}_{\rm fit}$ & $r_A^2\,[{\rm fm}^2]$ &
      $-2$LL & ${\cal N}_{\rm fit}$ & $r_A^2\,[{\rm fm}^2]$ &
      $-2$LL & ${\cal N}_{\rm fit}$ & $r_A^2\,[{\rm fm}^2]$ &
      $-2$LL & ${\cal N}_{\rm fit}$ & $r_A^2\,[{\rm fm}^2]$ &
      $N_{\rm bins}$ 
      \\ 
      \hline
      BNL1981  & 60.7 & $1.25^{+0.21}_{-0.14}$ & $0.61(13)$
               & 62.4 & $1.28^{+0.20}_{-0.17}$ & $0.83(24)$ 
               & 61.5 & $1.26^{+0.21}_{-0.18}$ & $0.74(25)$ 
               & 60.9 & $1.25^{+0.23}_{-0.19}$ & $0.67(24)$ & 47 \\
      ANL1982  & 43.2 & $1.40^{+0.25}_{-0.38}$ & $1.45^{+0.92}_{-0.49}$ 
               & 45.8 & $1.32^{+0.21}_{-0.18}$ & $1.04(24)$ 
               & 45.8 & $1.32^{+0.23}_{-0.20}$ & $1.03(25)$
               & 45.8 & $1.32^{+0.25}_{-0.21}$ & $1.05(24)$ & 46 \\
      FNAL1983 & 38.2 & $1.16^{+0.22}_{-0.16}$ & $0.33(7)$ 
               & 39.1 & $1.22^{+0.31}_{-0.25}$ & $0.64(31)$ 
               & 39.1 & $1.22^{+0.32}_{-0.25}$ & $0.63(30)$
               & 39.0 & $1.21^{+0.34}_{-0.26}$ & $0.60(35)$ & 28 
    \end{tabular}
  \end{ruledtabular}
\end{table*}

\begin{figure}[h!]
  \includegraphics[width=0.48\textwidth]{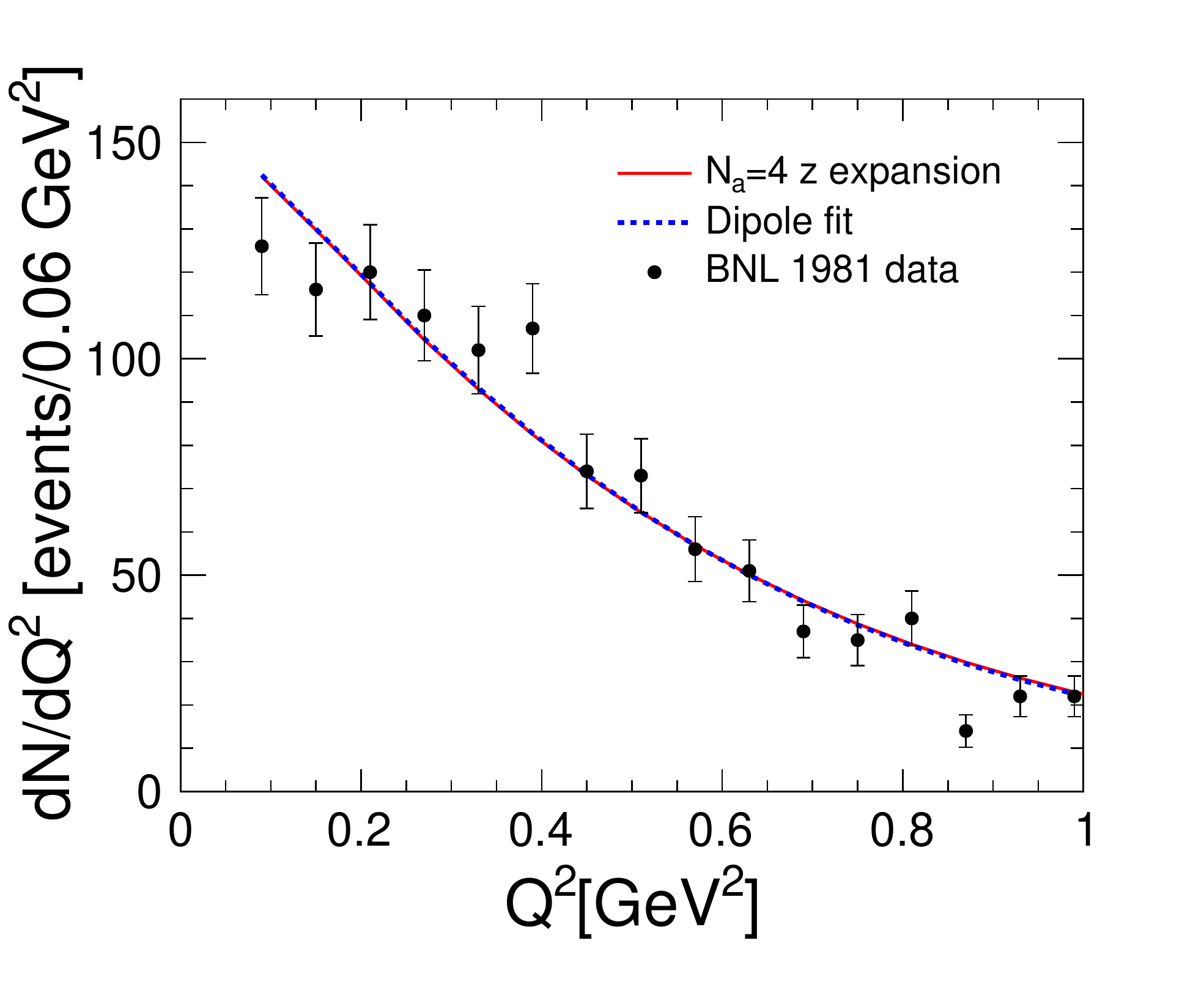}
  \includegraphics[width=0.48\textwidth]{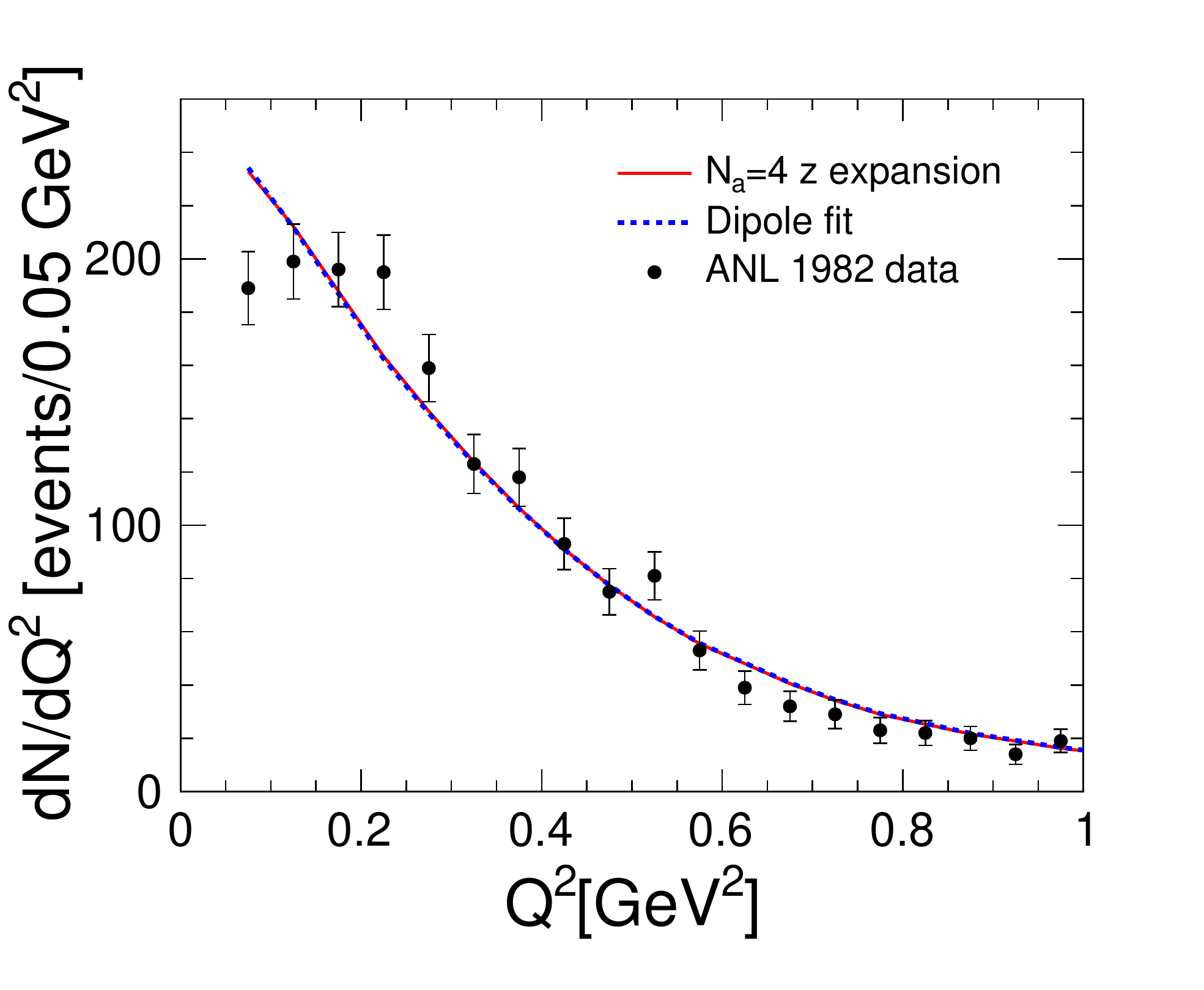}    
  \includegraphics[width=0.48\textwidth]{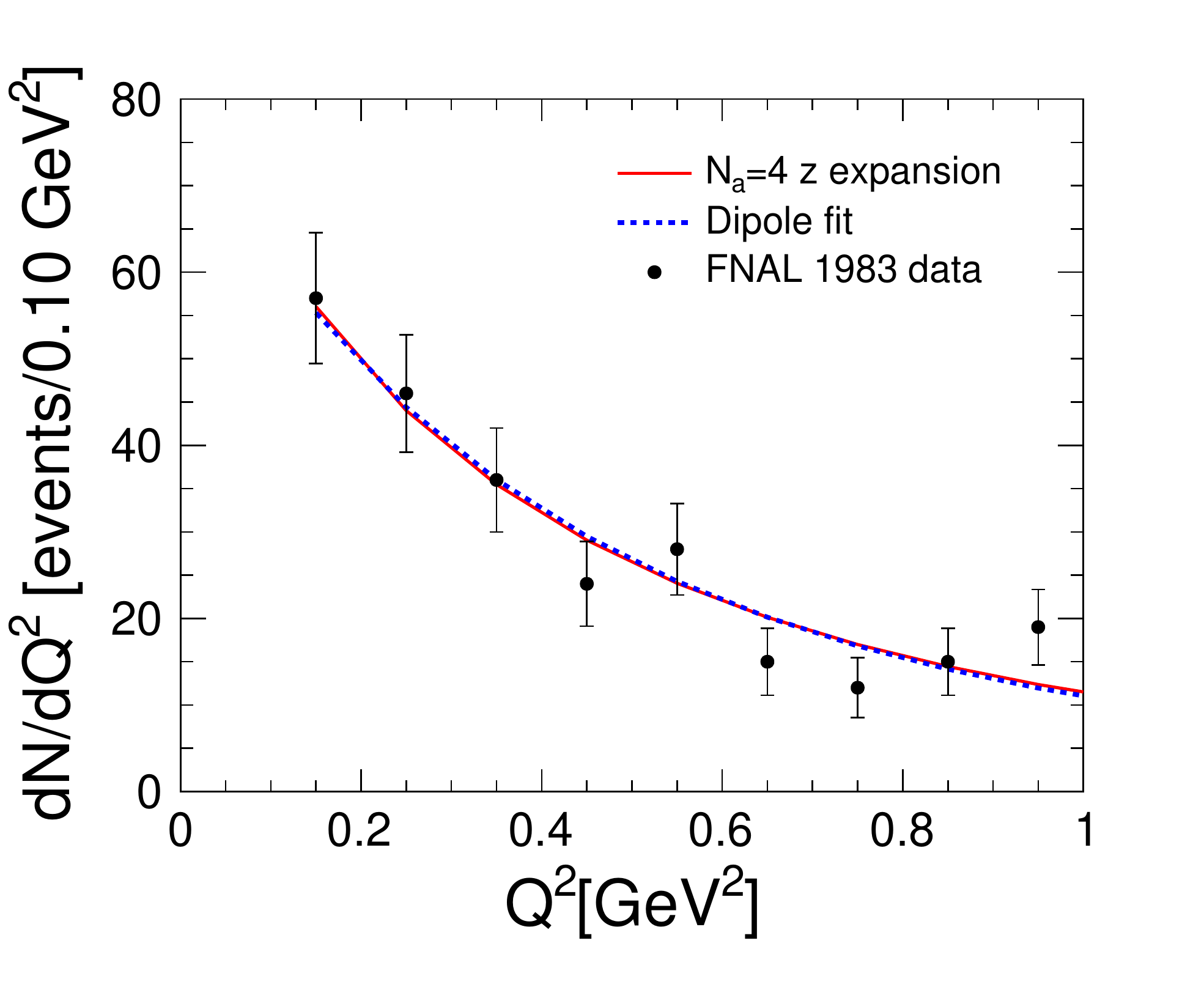}
  \caption{\label{fig:Q2_1}  Same as Fig.~\ref{fig:Q2}, but with $Q^2\le 1\,{\rm GeV}^2$.
    These fits correspond to the $N_a=4$ $z$ expansion in Table~\ref{tab:zfit1}. 
  }
\end{figure}

First, consider the removal of high $Q^2$ data, fitting to bins whose
center is within the restricted range $Q^2 \le 1\,{\rm GeV}^2$.  The
analog of Table~\ref{tab:zfit} for this case is given by
Table~\ref{tab:zfit1}.  Figure~\ref{fig:Q2_1} shows comparisons of
best fit curves and data points.  The analog of Eq.(\ref{eq:ak}) is
\begin{align}\label{eq:ak1}
  &[a_1, a_2, a_3, a_4 ] \big|_{Q^2\le 1\,{\rm GeV}^2} \nl
  &=\left\{ \begin{array}{lr}
    $[1.99(15), 0.5(1.1), -3.6(2.6), 1.1(2.7)]$ & ({\rm BNL})
    \\
    $[2.29(14), 0.2(0.9), -5.2(2.5), 2.9(2.7)]$ & ({\rm ANL})
    \\
    $[1.88(25), -0.9(1.6), -0.3(2.9), -0.3(2.8)]$ & ({\rm FNAL}) 
  \end{array}
  \right. \,.
  \end{align}

The omission of low-$Q^2$ data has a similarly large effect on the  fit
parameters.  Fitting to the range $Q^2 \ge 0.2\,{\rm GeV}^2$, the
results are given in Table~\ref{tab:zfit3}.  The $z$ expansion
coefficients are determined for $N_a=4$ to be
\begin{align}\label{eq:ak3}
  &[a_1, a_2, a_3, a_4 ] \big|_{Q^2\ge 0.2\,{\rm GeV}^2} \nl
  &=\left\{ \begin{array}{lr}
    $[2.35(10),\,-2.0(1.2),\,-1.4(2.8),\,1.4(2.7)]$ & ({\rm BNL})
    \\
    $[2.34(10),\,-3.6(1.2),\, 1.6(2.8),\,0.9(2.8)]$ &
    ({\rm ANL})
    \\
    $[2.04(16),\,-1.3(1.6),\,-0.5(3.0),\,0.1(2.8)]$ & ({\rm FNAL})
  \end{array}
  \right. \,.
\end{align}

Comparing the results in  Tables~\ref{tab:zfit}, \ref{tab:zfit1}, and
\ref{tab:zfit3}  and in Eqs.~(\ref{eq:ak}), (\ref{eq:ak1}), and
(\ref{eq:ak3}), we see that the leading $a_1$ and $a_2$ parameters
shift in some cases by about twice the statistical uncertainty of the
fits.  This reflects how different parts of the $Q^2$ range contribute
to tensions in the fit.   The minimum value of $-2{\rm LL} \sim
\chi^2$ decreases in both cases, closer to a range that would be
considered an adequate description of the data.  The improvement when
eliminating the low-$Q^2$ region is especially striking considering it
amounts to only two or three bins of data in each data set.

One method to translate the tensions in the fit to an uncertainty on
the fit parameters is to consider what additional error is necessary
to obtain a reduced $\chi^2$ of unity.  We include an error for each
data point proportional to the number of events in the original
$dN/dQ^2$ distribution.  This requires the use of a $\chi^2$
calculation instead of a log-likelihood fit, which we achieve by
limiting the test to the sample with $Q^2\le 1\,{\rm GeV}^2$.   Adding
this error in quadrature to the statistical error, we see that for
BNL, an additional $10\%$ error is required, while ANL requires an
additional $7.5\%$ error.  

\section{Systematic errors\label{sec:exptsyst}} 

The experimental uncertainties in the fits summarized in Table~\ref{tab:zfit}
correspond only to statistical errors on the number of events in each bin. 
With a framework in place to quantify theoretical form factor shape uncertainty, 
let us examine several sources of systematic error, and
their impact on the extraction of $F_A$.

Experimental systematic uncertainties come from the construction of
the neutrino flux, and from acceptance corrections.
A theoretical systematic error arises from uncertain modeling
of deuteron effects.

\subsection{Flux \label{sec:flux}}

Our procedure includes a self-consistent determination of the neutrino
flux for fits to the $Q^2$ distributions, as described in
Sec.~\ref{sec:Enu}.   Systematic uncertainty estimates in the
experimenter's {\it ab initio} flux do not apply.  Instead we check for
sensitivity to fluctuations in the number of events by varying one
$dN/dE_\nu$ bin by its statistical error, reextracting fit parameters,
and then repeating for all bins.  Adding errors in quadrature, the
result for the BNL data set is
\begin{align} \label{eq:vary}
\bar{a}_1 = 2.24 \pm 0.10_{{\rm stat.} Q^2} \pm 0.04_{{\rm stat.}
  E_\nu } \quad (\rm BNL1981) \,. 
\end{align}
Such an additional flux error is numerically subleading compared to
statistical error, and also to the systematic error assigned below to
account for deuteron and acceptance corrections.  We neglect it in our
final fits. 

The consistency of the flux procedure could also be impacted by
distortions of the $dN/dE_\nu$ distribution by $Q^2_{\rm min}$ cuts or
deuteron corrections.   Recall that the energy distribution from
BNL1981, but not from ANL1982 or FNAL1983, was corrected for these
effects.   We have checked that the resulting variations are even
smaller than the statistical fluctuations in Eq.~(\ref{eq:vary}), and
are neglected.

\subsection{Acceptance corrections} 

One source of uncertainty,  especially in the limit of very low $Q^2$,
is the acceptance corrections associated with human-eye scanning of
the bubble chamber photographs.  For example, Fig.~1 of ANL
1982~\cite{Miller:1982qi} provides an estimate of the scanning
efficiency ranging from $e= 90\pm 7\%$ at $0.05\,{\rm GeV}^2 < Q^2 <
0.1\,{\rm GeV}^2$ to $e= 98\pm 1\%$ for $Q^2 > 0.15\,{\rm GeV}^2$.  We
include a possible correlated efficiency correction by making the
following replacement in the efficiency-corrected number of events: 
\begin{align} \label{eq:eff}
   {dN \over e(Q^2)}  \to  {dN \over e(Q^2) + \eta \, de(Q^2)} = {dN\over
     e(Q^2)} \left( 1 + \eta {de(Q^2)\over e(Q^2)} \right)^{-1} \,. 
\end{align}
Here $\eta = 0\pm 1$ is a parameter introduced in the fit, and we use
a simple linear interpolation of the function in
Ref.~\cite{Miller:1982qi} for the efficiency $e(Q^2)$ and efficiency
error $de(Q^2)$.

In the BNL data set, an efficiency effect with similar magnitude is
presented, but not directly in the $Q^2$ variable.  For simplicity we
take the ANL function to represent possible effects also in the  BNL
and FNAL data sets, with independent floating scale parameters $\eta =
0\pm 1$ in Eq.~(\ref{eq:eff}).  The shape parameters and minimum
$-2{\rm LL}$ values are as follows, comparing results with and without
the acceptance correction, 
\begin{align}
  {\rm BNL}:\quad [\bar{a}_1, \, -2{\rm LL}]
  &= \left\{
  \begin{array}{ll}
    $[1.99(15),\,27.0]$
    & ( {\rm without} )
    \\
    $[2.04(15),\,26.0]$
    & ( {\rm with} )
  \end{array}\right. \,,
  \nl
  {\rm ANL}:\quad [\bar{a}_1, \, -2{\rm LL}]
  &= \left\{
  \begin{array}{ll}
    $[2.29(14),\,30.5]$
    & ( {\rm without} )
    \\
    $[2.38(14),\,26.3]$
    & ( {\rm with} )
  \end{array}\right. \,,
  \nl
  {\rm FNAL}:\quad [\bar{a}_1, \, -2{\rm LL}]
  &= \left\{
  \begin{array}{ll}
    $[1.88(25),\,8.2]$
    & ( {\rm without} )
    \\
    $[1.88(25),\,8.2]$
    & ( {\rm with} )
  \end{array}\right. \,.  
\end{align}
The parameter $\eta$ takes on values of $-1.9$, $-1.0$, and $+0.01$ for data
from ANL1982, BNL1981, and FNAL1983 respectively;  the negative values
indicate a pull to decrease the predicted cross section to match the data.
In each case there is only modest improvement in the fit quality,
and small impact on the form factor shape.   Acceptance corrections
within the quoted range have only minor impact. 

\subsection{Deuteron corrections}

\begin{figure}[tb]
\begin{center}
\includegraphics[width=0.49\textwidth]{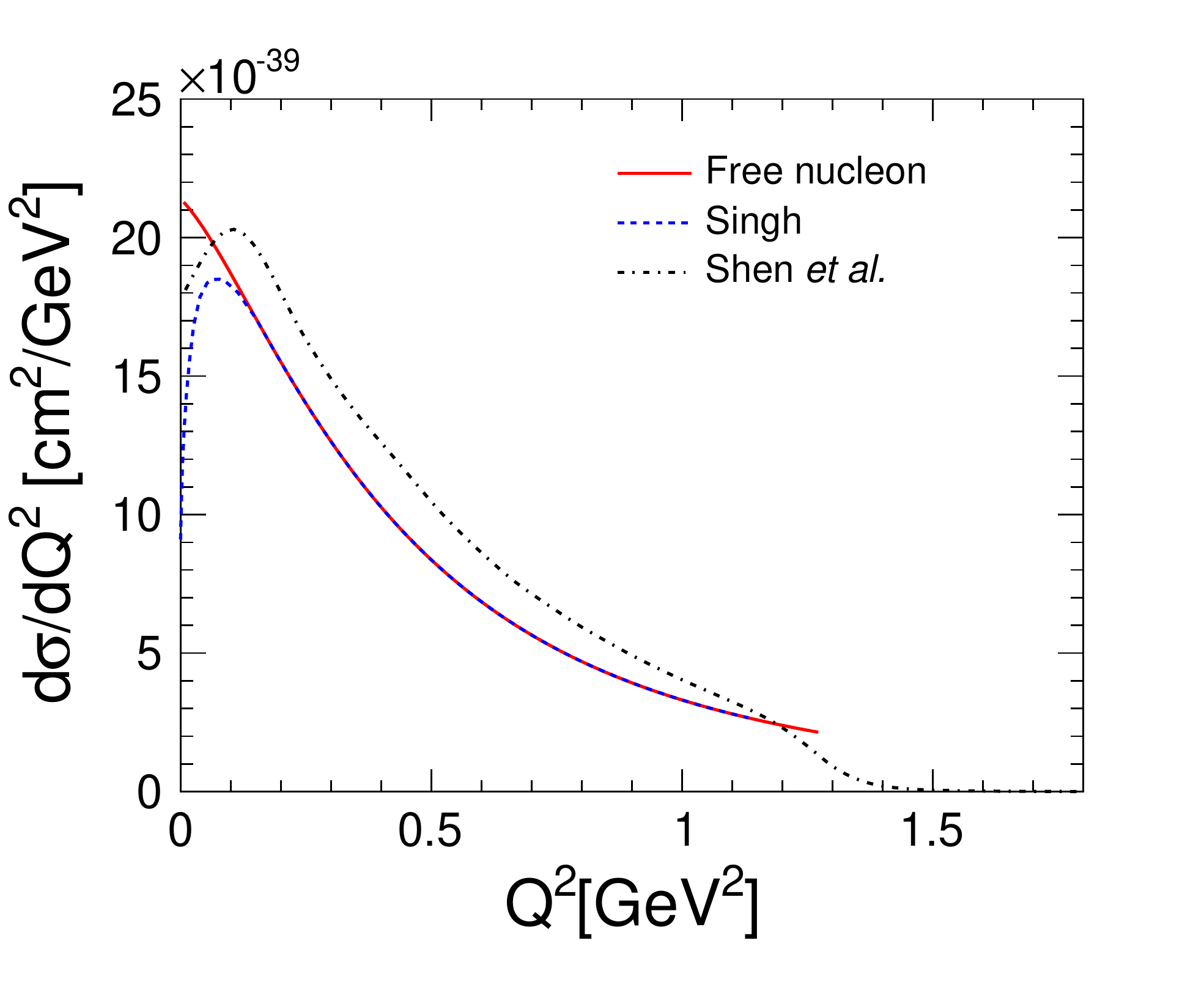}
\caption{
  Differential scattering cross sections for neutrino-deuteron scattering
  at $1\,{\rm GeV}$ neutrino energy, 
  employing different nuclear models.
  The solid (red) curve is the free-neutron result.   
  The dashed (blue) curve is obtained from the free-neutron result using
  the model from Ref.~\cite{Singh:1971md}, as in the original deuterium analyses. 
  The top dot-dashed (black) curve is extracted at $E_\nu=1\,{\rm GeV}$
  from Ref.~\cite{Shen:2012xz}.  The charged lepton mass is neglected in this plot. 
\label{fig:nuclear}
}
\end{center}
\end{figure}

The analysis to this point, like the original analyses, used the
deuteron correction model $R(Q^2)$ of Singh~\cite{Singh:1971md}.  This
model yields a suppression of the cross section for
$Q^2<0.16$~GeV$^2$.%
\footnote{  A follow-up analysis~\cite{Singh:1986xh} considers effects
  of meson exchange currents and alternate deuteron wave functions,
  with a total result very similar to Ref.~\cite{Singh:1971md}.  }
An example of a modern calculation with extended range in energy and
$Q^2$ is given by Shen {\it et al}. in Ref.~\cite{Shen:2012xz}.%
\footnote{See also Ref.~\cite{Moreno:2015nsa}.}
The Shen {\it et al}. model is overlaid with the original Singh model as
well as the free neutron model in Fig.~\ref{fig:nuclear}.  The Shen {\it et
al}. model deviates substantially from the free-neutron result at the
$\sim 20\%$ level over a broad $Q^2$ range.  These models do not
constitute an estimate of the uncertainty on deuteron corrections, but
suggest an avenue for future work even if there are no future
measurements on deuterium.

Assuming an energy independent, but $Q^2$ dependent, deuteron
correction, the change in the fit results can be compared.  For
illustration, we employ the results of Ref.~\cite{Shen:2012xz} at
$E_\nu=1\,{\rm GeV}$, and  limit attention to $Q^2\le 1\,{\rm GeV}^2$,
i.e.,  the configuration of Table~\ref{tab:zfit1} and
Eq.~(\ref{eq:ak1}).  Shape parameter and minimum $-2{\rm LL}$ values
are 
\begin{align}
  {\rm BNL}:\quad [\bar{a}_1, \, -2{\rm LL}]
  &= \left\{
  \begin{array}{ll}
    $[1.99(15),\,27.0]$
    & ( {\rm Singh} )
    \\
    $[2.16(14),\,25.1]$
    & ( {\rm Shen\,et\,al.} )
  \end{array}\right. \,,
  \nl
  {\rm ANL}:\quad [\bar{a}_1, \, -2{\rm LL}]
  &= \left\{
  \begin{array}{ll}
    $[2.29(14),\,30.5]$
    & ( {\rm Singh} )
    \\
    $[2.46(13),\,29.2]$
    & ( {\rm Shen\,et\,al.} )
  \end{array}\right. \,,
  \nl
  {\rm FNAL}:\quad [\bar{a}_1, \, -2{\rm LL}]
  &= \left\{
  \begin{array}{ll}
    $[1.88(25),\, 8.2]$
    & ( {\rm Singh} )
    \\
    $[2.00(25),\, 9.1]$
    & ( {\rm Shen\,et\,al.} )
  \end{array}\right. \,.  
\end{align}
The extracted form factor shifts to mimic the difference in the curves
in Fig.~\ref{fig:nuclear}, and there is slight improvement in fit
quality for two of the three data sets. 

\subsection{Final systematic error budget \label{sec:systfinal}}

The most important systematic uncertainties are the two that
significantly modify the $Q^2$ distribution:  acceptance corrections
and the deuteron correction.  In our final analysis, we modify the
original fits displayed in Table~\ref{tab:zfit1}.  First, we allow a
correlated acceptance correction as in Eq.~(\ref{eq:eff}).  Second, we
include a 10\% error added in quadrature to statistical error in each
$Q^2$ bin to account for residual deuteron or other systematic
corrections, as described at the end of Sec.~\ref{sec:range}.  With
these corrections in place, we perform a $\chi^2$ fit to all data up
to $Q^2=1\,{\rm GeV}^2$.  The neglect of data above $Q^2=1\,{\rm
  GeV}^2$ has only minor impact on the extraction of $F_A(q^2)$,
and allows a simple treatment of these combined uncertainties with
full covariance using a $\chi^2$ fit.  

As an alternative, we also provide a log-likelihood fit to the data up
to $Q^2=3\,{\rm GeV}^2$, but without inflated errors to account for
deuterium and other residual systematics.  This has the benefit of
including data over the entire kinematic range, but omits sources 
of systematic error that would need to be treated separately.

\section{Axial form factor extraction\label{sec:extract}}

The best axial form factor is extracted from a joint fit to the three data sets.
We choose $N_a=4$ free parameters with $t_{0} = t_{0}^{\text{optimal}}(1\,{\rm GeV}^{2})$
and data with $Q^2 \le 1\,{\rm GeV}^2$.
As discussed above, this corresponds to a $k_{\rm max}=8$ $z$ expansion,
where five linear combinations of coefficients are fixed by the $Q^2=0$
constraint and by the four sum rules (\ref{eq:sumrules}).  The
acceptance correction free parameter is independent for each
experiment in the joint fit.

\begin{figure}[ht]
\begin{center}
  \includegraphics[width=0.49\textwidth]{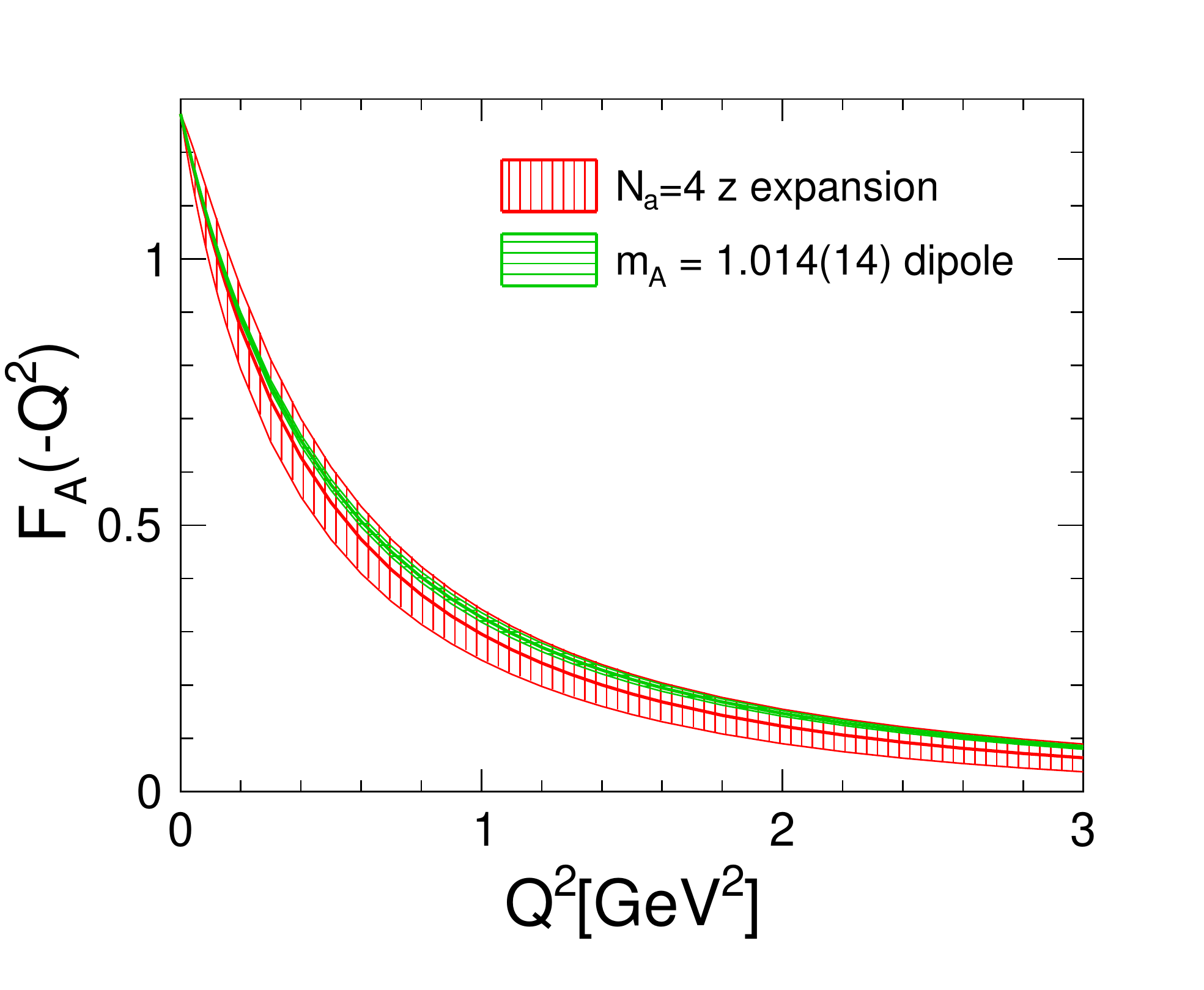}
  \includegraphics[width=0.49\textwidth]{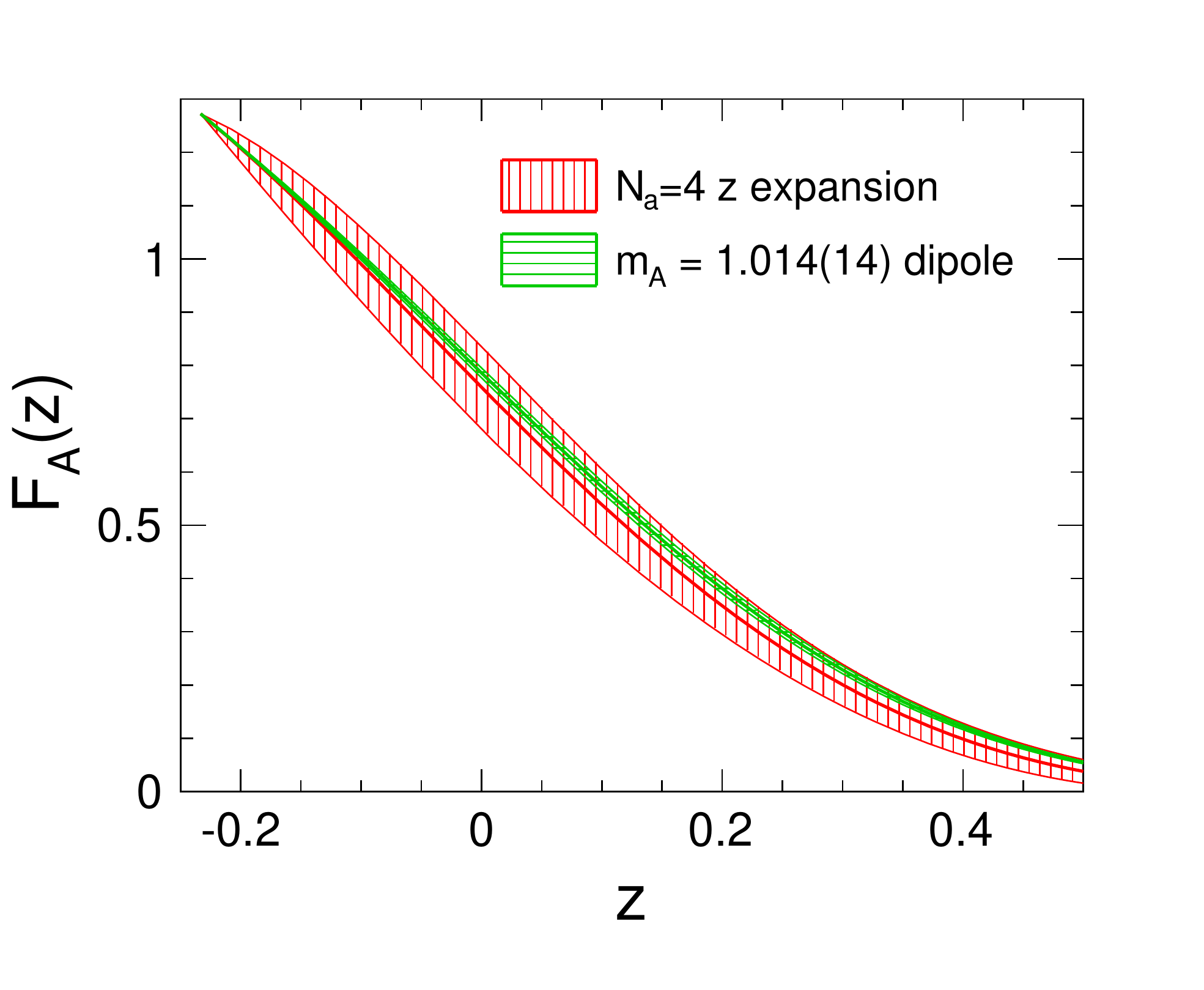}
\caption{\label{fig:FA}
  Final form factor from Eqs.~(\ref{eq:ff}), (\ref{eq:covdiag}) and
  (\ref{eq:fferr}).
  Also shown is the dipole axial form factor with axial mass
  $m_A=1.014(14)~{\rm GeV}$~\cite{Bodek:2008epjc}.
}
\end{center}
\end{figure}

Our knowledge of the axial form factor resulting from deuterium
scattering data is summarized by constraints on the coefficients
$a_k$.   Central values and $1\sigma$ errors determined from
$\Delta\chi^2=1$ are%
\footnote{ The complete specification for the form factor involves the
  normalization $g_A=-1.2723$ from Table~\ref{tab:constants}; the pion
  mass $m_\pi=0.14\,{\rm GeV}$ employed in the specification of
  $t_{\rm cut}=9m_\pi^2$ in Eq.~(\ref{eq:zdef}); and the choice
  $t_0=-0.28\,{\rm GeV}^2$.  The remaining coefficients, $a_0$, $a_5$,
  $a_6$, $a_7$ and $a_8$, are determined by $F_A(0)=g_A$, and by the
  sum rule constraints (\ref{eq:sumrules}); for ease of comparison we
  list the complete list of central values here:
  $[a_0,\cdots,a_8]=[-0.759, 2.30, -0.6, -3.8, 2.3, 2.16, -0.896,
    -1.58, 0.823]$.}
\begin{align}
\label{eq:ff}
[a_1, a_2, a_3, a_4] = [ 2.30(13), -0.6(1.0), -3.8(2.5), 2.3(2.7) ]
\,.
\end{align}
The diagonal entries of the error (covariance) matrix, computed from
the inverse of the Hessian matrix for $\chi^2(\{a_k\})$, are 
\begin{align} \label{eq:covdiag}
E_{\rm diag.} = [0.0154, 1.08, 6.54, 7.40] \,. 
\end{align}
Note that $(E_{\rm diag.})_i \approx (\delta a_i)^2$, reflecting
approximately Gaussian behavior.  The four-dimensional correlation
matrix is
\begin{align} \label{eq:fferr}
C_{ij} = \left( \begin{array}{cccc} 1 & 0.350  & -0.678 &  0.611
  \\ 0.350 & 1  & -0.898 &  0.367 \\ -0.678 & -0.898 & 1  & -0.685
  \\ 0.611 &  0.367 & -0.685  & 1 \\
  \end{array} \right) \,. 
\end{align}
and as usual the error matrix is given by $E_{ij} = \delta a_i \delta
a_j C_{ij}$.   This description can be systematically improved when
and if further data or externally constrained deuterium models become
available.  The form factor is plotted versus $Q^2$ and versus $z$ in
Fig.~\ref{fig:FA}, and compared with a previous
world average dipole form factor from Ref.~\cite{Bodek:2008epjc}.

We also provide an alternate log-likelihood determination of the axial form factor
to the range $Q^2<3.0~{\rm GeV}^2$, but without deuteron systematic corrections.
Central values and $1\sigma$ errors determined from $\Delta(-2LL)=1$
are 
\begin{align}
\label{eq:ff2}
  [a_1, a_2, a_3, a_4] =  [2.28(8), 0.25(95), -5.2(2.3), 2.6(2.7)] \,.
\end{align}
The diagonal entries of the error matrix are
\begin{align} \label{eq:covdiag2}
E_{\rm diag} = [0.00635, 0.781, 4.49, 6.87] \,,
\end{align}
and the four-dimensional correlation matrix is
\begin{align} \label{eq:fferr2}
C_{ij} = \left( \begin{array}{cccc} 1 & 0.321  & -0.677 &  0.761
  \\ 0.321 & 1  & -0.889 &  0.313 \\ -0.677 & -0.889 & 1  & -0.689
  \\ 0.761 &  0.313 & -0.689  & 1 \\
  \end{array} \right) \,. 
\end{align}

\section{Applications \label{sec:nuclear}}

Having presented the axial form factor with errors and correlations
amongst the coefficients, we may systematically  compute derived
observables that depend on this function.  We consider several
applications of our results.

\subsection{Axial radius}  

\begin{table}[t]
  \caption{\label{tab:rA}
    Axial radius extracted using best values from Table~\ref{tab:constants}, and default priors
    as discussed in the text. Note that the joint fit is not an average, but a simultaneous
    fit to all of the data sets.
  }
    \begin{ruledtabular}
    \begin{tabular}{llll}
      Data set & $r_A^2\,[{\rm fm}^2]$  & $r_A^2\,[{\rm fm}^2]$ &
      $r_A^2\,[{\rm fm}^2]$ \\
& ($N_a=3$) & ($N_a=4$) & ($N_a=5$) \\
      \hline
      BNL 1981 & $0.56(23)$ & $0.52(25)$ &
      $0.48(26)$
      \\
      ANL 1982 & $0.69(21)$ & $0.63(23)$ &
      $0.57(24)$ 
      \\
      FNAL 1983 & $0.63(34)$ & $0.64(35)$ &
      $0.64(35)$ 
      \\
      Joint Fit & $0.54(20)$ & $0.46(22)$ &
      $0.39(23)$ 
    \end{tabular}
  \end{ruledtabular}
\end{table}

We begin with the axial radius, defined in Eq.~(\ref{eq:rA}). 
While the radius by itself is not the only quantity of interest
to neutrino scattering observables, it is only through the $q^2\to 0$
limit that a robust comparison can be made to other processes such as
pion electroproduction.

The form factor coefficients and error matrix from the $\chi^2$ fit in
Sec.~\ref{sec:extract} determine the radius as 
\begin{align} \label{eq:rAfinal}
r_A^2 = 0.46(22)\,{\rm fm}^2 \,. 
\end{align}
The constraint is much looser than would be obtained by restricting to
the  dipole model, cf. Table~\ref{tab:zfit}.%
\footnote{Extractions of the radius from electroproduction data are
  also strongly influenced by the dipole
  assumption~\cite{Bhattacharya:2011ah}.}
For comparison, let us consider the constraints from individual
experiments.    Table~\ref{tab:rA} gives results for $N_a=3,4,5$ free
parameters, with errors determined from the error matrix in
Eqs.~(\ref{eq:covdiag}) and (\ref{eq:fferr}). The
results from individual experiments are consistent with the joint fit.
Note that the joint fit is not simply the average of the individual
fits.   This situation arises from a slight tension between data and
Gaussian coefficient constraints (\ref{eq:bound1}) when comparing a single data set to
the statistically more powerful combined data.

\subsection{Neutrino-nucleon quasielastic cross sections}

\begin{figure}[tb]
\begin{center}
\includegraphics[width=0.49\textwidth]{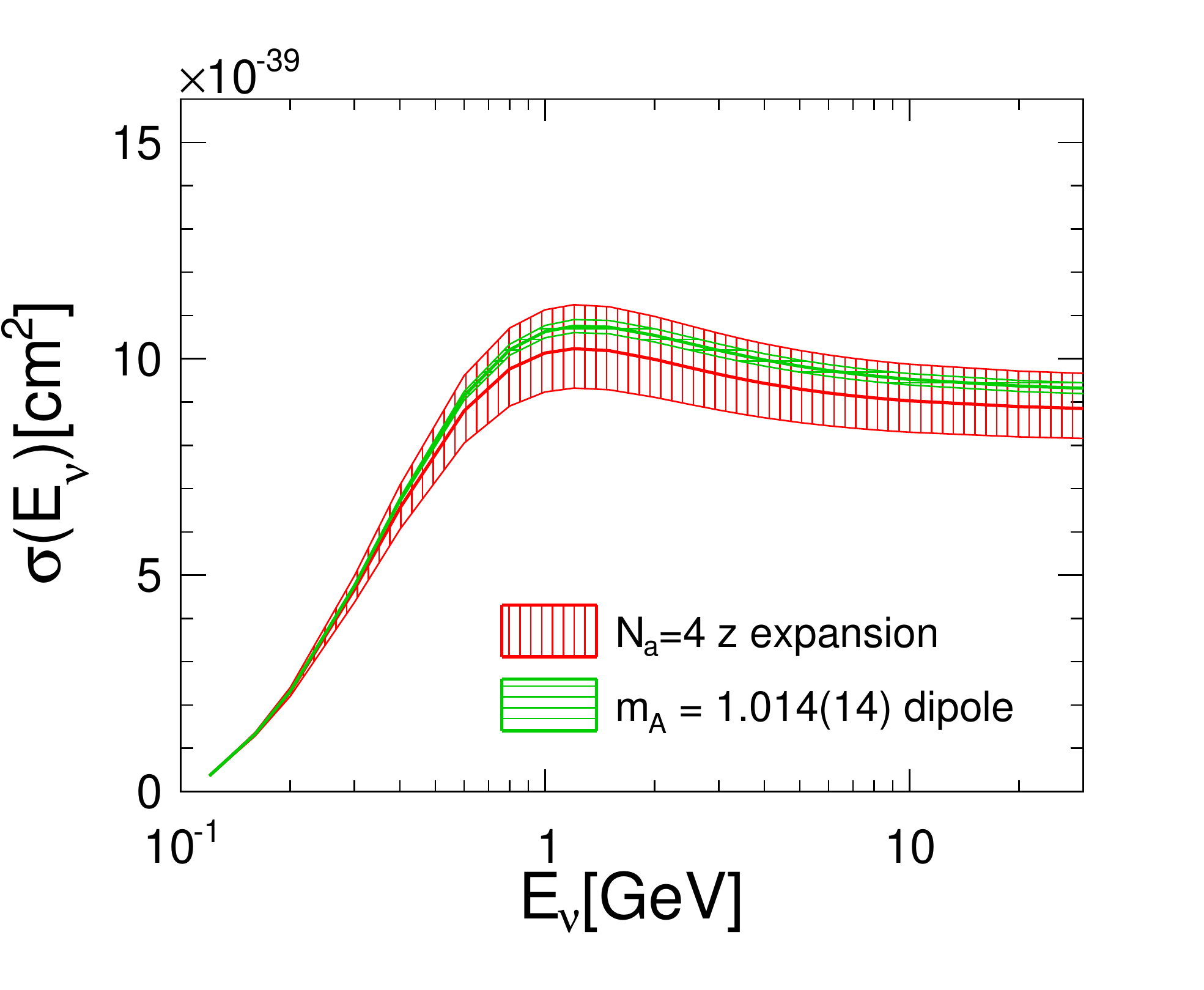}
\includegraphics[width=0.49\textwidth]{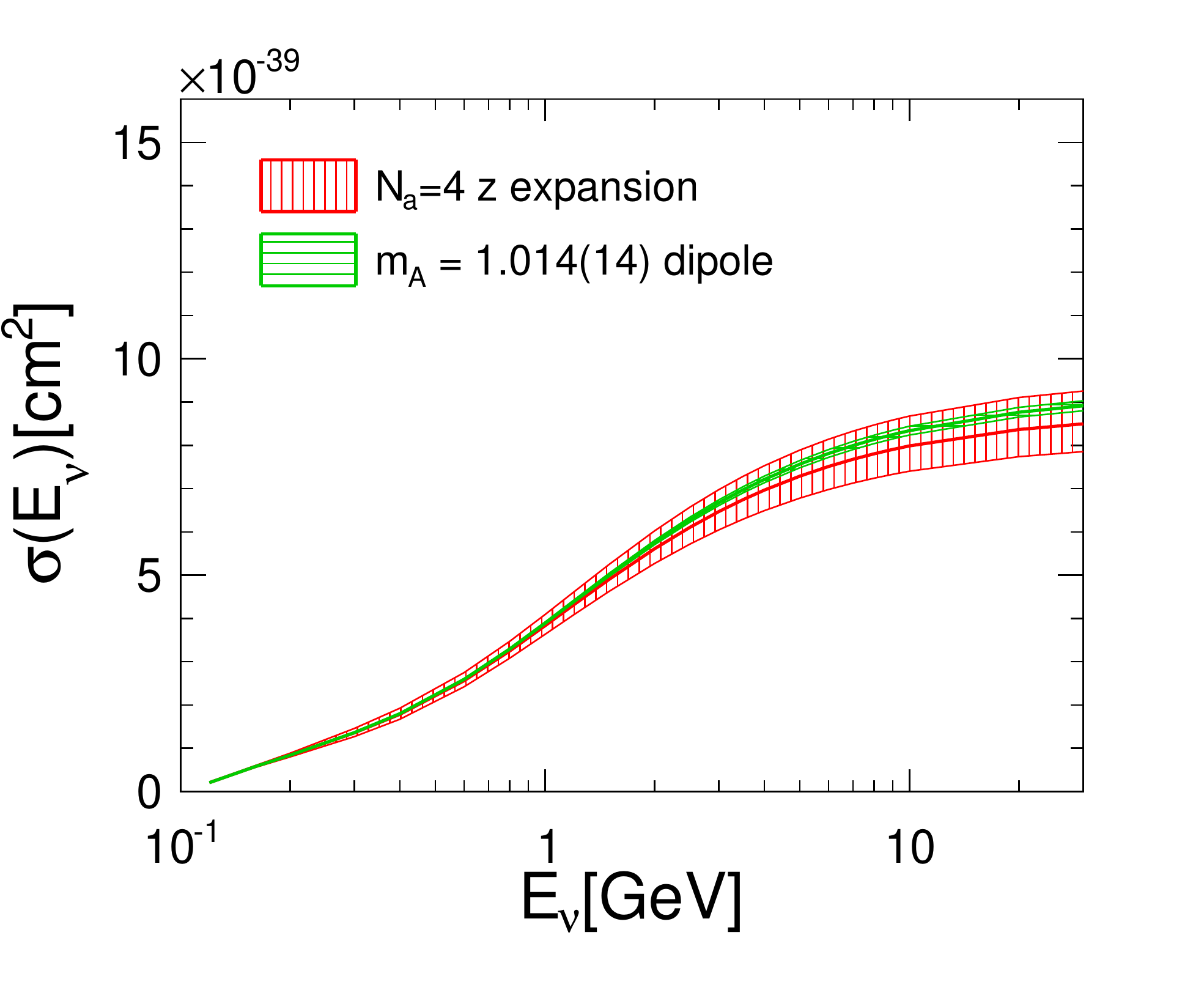}
\caption{  \label{fig:freeCCQE}
  Free nucleon CCQE cross section computed from Eqs.~(\ref{eq:ff}), (\ref{eq:covdiag}) and
  (\ref{eq:fferr}), for neutrino-neutron (top) and
  antineutrino-proton (bottom) scattering. 
  Also shown are results using dipole axial form factor with axial mass
  $m_A=1.014(14)~{\rm GeV}$~\cite{Bodek:2008epjc}.
}
\end{center}
\end{figure}

Current and future neutrino oscillation experiments will precisely
measure neutrino mixing parameters, determine the neutrino mass
hierarchy, and search for possible CP violation and other new
phenomena.  This program relies on accurate predictions, with
quantifiable uncertainties, for neutrino interaction cross sections.
As the simplest examples, consider the charged-current quasielastic
cross section $\sigma(E_\nu)$ for neutrino (antineutrino) scattering
on an isolated neutron (proton).  

The best fit cross section and uncertainty are shown in
Fig.~\ref{fig:freeCCQE}, and compared to the prediction of dipole
$F_A$ with axial mass $m_A=1.014(14)$~\cite{Bodek:2008epjc}.  At
representative energies, the cross sections and uncertainties shown in
Fig.~\ref{fig:freeCCQE} are 
\begin{align}\label{eq:nun}
  \sigma_{\nu n \to \mu p}(E_\nu=1\,{\rm GeV}) &= 10.1(0.9)\times 10^{-39}\,{\rm cm}^2\,, \nl
  \sigma_{\nu n \to \mu p}(E_\nu=3\,{\rm GeV}) &= 9.6(0.9)\times 10^{-39}\,{\rm cm}^2\,, 
\end{align}
for neutrinos and
\begin{align}\label{eq:nup}
  \sigma_{\bar{\nu} p \to \mu n}(E_\nu=1\,{\rm GeV}) &= 3.83(23)\times 10^{-39}\,{\rm cm}^2\,, \nl
  \sigma_{\bar{\nu} p \to \mu n}(E_\nu=3\,{\rm GeV}) &= 6.47(47)\times 10^{-39}\,{\rm cm}^2\,, 
\end{align}
for antineutrinos. 

\subsection{Neutrino nucleus cross sections} 

\begin{figure}[t]
\begin{center}
\includegraphics[width=0.49\textwidth]{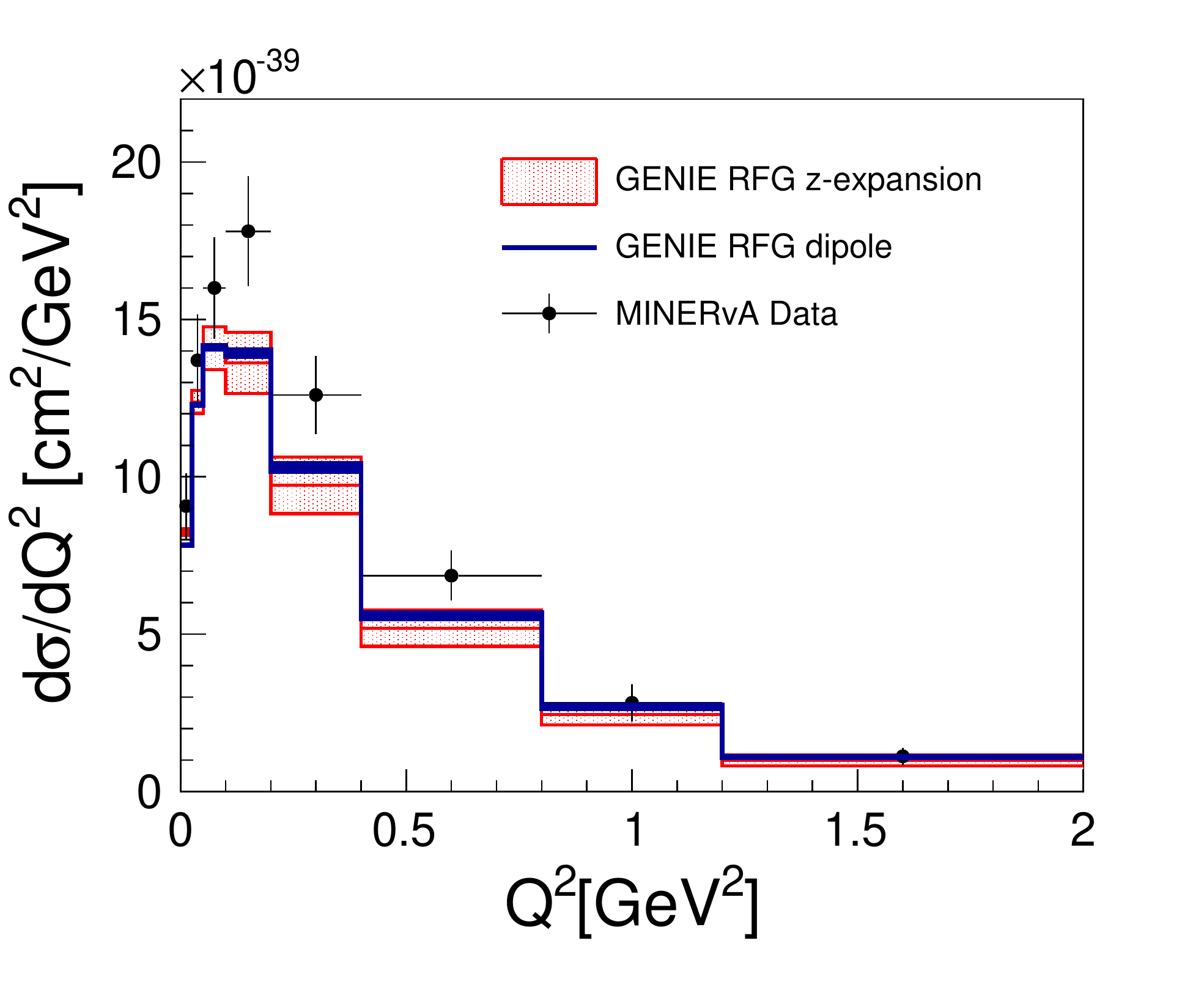}
\caption{
  Cross section for charged-current quasielastic events from the
  MINERvA experiment~\cite{Fiorentini:2013ezn}
  as a function of reconstructed $Q^2$, compared with prediction using relativistic Fermi gas (RFG)
  nuclear model with $z$ expansion axial form factor extracted from deuterium data. 
  MINERvA data uses an updated flux prediction from~\cite{Betancourt:2015nuint}.
  Also shown are results using the same nuclear model but dipole form factor with axial mass
  $m_A=1.014(14)~{\rm GeV}$~\cite{Bodek:2008epjc}.
  \label{fig:minerva}
}
\end{center}
\end{figure}

Connecting nucleon-level information to experimentally observed
neutrino-nucleus scattering cross sections requires data-driven
modeling of nuclear effects.  Our description of the axial form factor
and uncertainty in Eqs.~(\ref{eq:ff}), (\ref{eq:covdiag}), and
(\ref{eq:fferr}) can be readily implemented in neutrino event
generators that interface with nuclear models.%
\footnote{The $z$ expansion will be available in GENIE production release v2.12.0.
  The code is currently available in the GENIE trunk prior to its official release.
  The module provides full generality of the $z$ expansion, and supports
  reweighting and error analysis with correlated parameters.}

A multitude of studies and comparisons are possible.  As illustration,
consider MINERvA quasielastic data on
carbon~\cite{Fiorentini:2013ezn}.  Figure~\ref{fig:minerva} shows a
comparison of the $Q^2$ distribution of measured events with the
predictions from our $F_A(q^2)$, using a relativistic Fermi gas
nuclear model in the default configuration of the GENIE v2.8 neutrino event
generator~\cite{Andreopoulos:2009rq}.  For comparison, we display the
result obtained using a dipole $F_A$ with axial mass central value and
error as quoted in the world average of Ref.~\cite{Bodek:2008epjc}.
The central curves differ in their kinematic dependence, and the
dipole result severely underestimates the uncertainty propagated from
deuterium data. 

The $z$ expansion implementation within GENIE includes a complete
description of parameter errors and correlations.  This will provide a
systematic approach for testing different nuclear models
and fitting nuclear model parameters, and for propagating
uncertainties in nucleon-level amplitudes  through to oscillation
observables. 

\subsection{Discussion}

The dipole ansatz has been commonly used to parametrize the axial
form factor in neutrino cross section predictions.  The axial mass
parameter in this ansatz often appears with either a very small
uncertainty, e.g. $m_A = 1.014(14)\,{\rm GeV}$~\cite{Bodek:2008epjc},
or a very large uncertainty, e.g. $m_A = 1.21(45)\,{\rm
  GeV}$~\cite{Abe:2015awa}.

In the first case, the small error estimate results from the
restrictive dipole ansatz, and is likely an underestimate of the
actual uncertainty:  as a point of comparison, the $\lesssim 1.5\,\%$
axial radius error is comparable to or smaller than the uncertainty on
the proton charge radius~\cite{Bernauer:2013tpr,Lee:2015jqa}.  Recall
that the charge radius is defined for the vector charge form factor
analogously to the axial radius for the axial form factor.  In
contrast to the axial radius from neutrino-deuteron scattering, the
charge radius from electron-proton scattering involves much {\it
  higher} statistics, a {\it monoenergetic} beam, and a {\it simpler},
proton, target.

In the second case, the large uncertainty on $m_A$ is typically
included to account for tensions in external inputs from other
experiments~\cite{Abe:2015awa}, and/or poorly constrained nuclear
effects.  Neither of these approaches is suited to the kinds of
analyses that can be  undertaken with modern cross section data such
as the MINERvA example considered in Fig.~\ref{fig:minerva}.
Underestimating nucleon-level uncertainties will bias conclusions
about neutrino parameters or nuclear models.   Inflating errors on
$m_A$ within a dipole ansatz fails to capture the correct kinematic
dependence of either nucleon-level uncertainties, or of nuclear
corrections%
\footnote{
Nondipole parametrizations have been considered in~\cite{Amaro:2015lga,Bodek:2007ym}. Similar
remarks apply to these examples.
}.

\section{Summary and conclusion \label{sec:conclusion}}

The constraints of elementary target data are critical to precision
neutrino-nucleus cross sections underlying the accelerator neutrino
program.  Oscillation experiments rely on event rate predictions using
nucleon-level amplitudes corrected for nuclear effects.  Cross section
experiments on nuclear targets can measure these nuclear effects but a
complete accounting of uncertainty in nucleon-level amplitudes is
critical for disentangling nucleon-level, nuclear-level, and flux
uncertainties, and for determining final sensitivity to fundamental
neutrino parameters. 

The axial form factor is a prominent source of nucleon-level
uncertainty.  We have analyzed the world data set for quasielastic
neutrino-deuteron scattering using a model-independent description
of the axial form factor.  Our final results are presented with
central values (\ref{eq:ff}), errors (\ref{eq:covdiag}) and
correlations (\ref{eq:fferr}).   Any observable depending on
the axial form factor may be computed from these results, with a
complete error budget. 

The axial radius, governing the shape of the axial form factor, 
is presented in Eq.~(\ref{eq:rAfinal}). It  has a significantly larger uncertainty
than previously estimated based on the unjustified dipole ansatz.
Benchmark total cross sections on nucleon targets are presented in
Fig.~\ref{fig:freeCCQE} and Eqs.~(\ref{eq:nun}) and~(\ref{eq:nup}).  The
incorporation of nuclear effects with the RFG model is illustrated in
Fig.~\ref{fig:minerva}. 

The form factor and uncertainty budget presented here are important
new inputs to the neutrino cross section effort.   It is interesting
to investigate potential impacts and interplay with a variety of other
processes such as neutrinoless double beta decay matrix
elements~\cite{Simkovic:2007vu,Holt:2013tda} and the muon capture rate
in muonic hydrogen~\cite{Andreev:2012fj}.  The methodology presented
can be revised or extended if new information becomes available.
Future hydrogen or deuterium data would be trivial to include.
Updated calculations for neutrino-deuteron scattering, especially if
accompanied by an uncertainty, can be readily incorporated on top of
this result.  Lattice QCD holds promise to determine the axial form
factor over much of the relevant $Q^2$ range, in a manner that is free
from nuclear corrections~\cite{Bazavov:2009bb,Dinter:2011plb,Bazavov:2013prd,Bhattacharya:2014prd,Green:2014plb,Bazavov:2015}. 

\vskip 0.2in
\noindent
{\bf Acknowledgments} We thank L.~Alvarez Ruso, J.~R. Arrington, H.~Budd, S.~Bacca,
A.~Kronfeld, T.~Mann, J.~Morfin, G.~Paz, and J.~W.~Van Orden for discussions,
and R.~Schiavilla for providing data files and interpretation of the results of
Ref.~\cite{Shen:2012xz}.  RG was supported by NSF Grant No. 1306944. 
Research of RJH and ASM was supported by
DOE Grant No. DE-FG02-13ER41958.  RG and RJH thank CETUP* (Center for
Theoretical Underground Physics and Related Areas), for its
hospitality and partial support during the 2014 Summer Program.
Research of ASM also supported by the U.S. Department of Energy, 
Office of Science Graduate Student Research (SCGSR) program.
The SCGSR program is administered by the Oak Ridge Institute
for Science and Education for the DOE under contract number DE-AC05-06OR23100.
\vskip 0.11in
\noindent

\end{document}